\documentclass[twocolumn]{aastex631}
\usepackage{amsmath,amssymb,bm,graphicx}
\providecommand{\uat}[2]{\href{http://astrothesaurus.org/uat/#2}{#1 (#2)}}
\newcommand{\X}{\boldsymbol{X}}
\newcommand{\R}{\boldsymbol{R}}
\newcommand{\K}{\boldsymbol{K}}
\newcommand{\avg}[1]{\left\langle #1 \right\rangle}
\newcommand{\Rea}{\operatorname{Re}}
\newcommand{\Ima}{\operatorname{Im}}
\shorttitle{Directions in Mixed Faraday Turbulence}
\shortauthors{Lazarian, Pogosyan, \& Melnichenka}
\begin{document}
\title{Directional Statistics of Synchrotron Polarization in Mixed Emitting and Faraday-Rotating Turbulence}

\author[0000-0002-7336-6674]{A. Lazarian}
\affiliation{Department of Astronomy, University of Wisconsin--Madison, Madison, WI 53706, USA}

\author[0000-0002-7998-6823]{D. Pogosyan}
\affiliation{Department of Physics, University of Alberta, Edmonton, AB T6G 2E1, Canada}

\author[0009-0004-1969-779X]{A. Melnichenka}
\affiliation{Department of Physics, Berea College, Berea, KY 40404, USA}

\begin{abstract}
Polarization directions provide a sensitive probe of Faraday rotation without weighting the signal by polarized brightness. We develop their spatial statistics for a turbulent volume in which synchrotron emission and Faraday rotation coexist. For independent, zero-mean Gaussian polarized emissivity, we derive an exact relation for the directional correlation. We derive strong stochastic coherence-depth scalings, including an additional interval for steep Faraday-density spectra, and distinguish them from the conditions needed to infer directional spectral slopes. We also quantify the weak-rotation response of a coherent uniform emitter and demonstrate that the scalings of the normalized tangential polarization gradient follow directly from the direction spectrum. A combined use of the techniques, including those introduced earlier in Lazarian \& Pogosyan (2016), provides a way for detailed study of astrophysical turbulence.

\end{abstract}

\section{Introduction}\label{sec:intro}
Magnetized turbulence controls the transport and evolution of the interstellar medium, affects cosmic-ray propagation, and shapes the structure of galactic and extragalactic radio emission \citep{ElmegreenScalo2004,BrandenburgLazarian2013,Akahori2018}. Radio observations offer complementary views of this turbulence. Synchrotron intensity depends primarily on the magnetic field perpendicular to the line of sight (LOS) and on the relativistic electrons. The polarization direction records the projected field geometry, whereas Faraday rotation probes the thermal-electron-weighted LOS field. Recovering their spatial statistics is particularly valuable because direct three-dimensional measurements of astrophysical magnetic fields are generally unavailable \citep{Haverkorn2015,Beck2015}.

The traditional tools include polarization-angle dispersion, rotation-measure structure functions, depolarization, and Faraday-depth reconstruction \citep{Burn1966,Tribble1991,Brentjens2005,Hildebrand2009,Houde2009}. RM synthesis exploits frequency coverage to separate components in Faraday depth, but an extended turbulent emitting region does not in general correspond to a set of geometrically isolated emitting layers. Likewise, the Stokes maps at one wavelength contain the cumulative effects of intrinsic polarization and propagation. A practical statistical description must distinguish the field responsible for synchrotron emission from the Faraday-rotating plasma, even when they occupy the same volume.

Considerable progress has been made in relating radio observables to the turbulent spectrum. \citet{LazarianPogosyan2012} (LP12) derived synchrotron-intensity correlations for magnetized turbulence, including anisotropy and dependence on the relativistic-electron index. \citet{LazarianPogosyan2016} (LP16) developed the spatial and wavelength statistics of complex polarization for both coextensive emitting/rotating media and foreground Faraday screens. Their polarization-frequency and wavelength-derivative statistics distinguish rotation by a coherent LOS field from fluctuations of the Faraday density. These methods have been tested in synthetic and MHD data \citep{Herron2016,Zhang2016,Lee2016,Zhang2018,Wang2022}, and subsequent work has explored additional spectral, interferometric, and higher-order signatures \citep{Zhang2020,ZhangXW2023,Wang2024,Xiao2025}. The strong-mean spatial approximation of LP16 requires a Fourier treatment rather than the previously assumed depth cutoff; we provide that correction in Section~\ref{sec:lp16correction}.

A complementary approach follows the spatial organization of polarization itself. Maps of Stokes-polarization gradients reveal fluctuations in magneto-ionic material \citep{Gaensler2011,Burkhart2012,Iacobelli2014,Herron2017}. Generalized directional derivatives and other invariant polarization diagnostics resolve the radial and tangential changes in the $Q$--$U$ plane \citep{Herron2018a,Herron2018b}. Synchrotron-intensity gradients and Faraday-sensitive polarization gradients have also been used to trace magnetic-field orientations and to explore the emitting and rotating regions at different effective depths \citep{LazarianEtAl2017SIG,LazarianYuen2018,Zhang2019,Wang2021}. These methods emphasize local gradients or anisotropy. A complementary statistic is the \emph{spatial spectrum of polarization directions}, introduced and numerically tested for a foreground Faraday screen by \citet{Melnichenka2026}. Here we study the same amplitude-normalized observable when the synchrotron emission and the Faraday rotation occur in the same turbulent volume.

We use the normalized Stokes field to represent the polarization direction, including its $180^\circ$ ambiguity:
\begin{equation}
 u(\X,\lambda^2)=\frac{Q+iU}{(Q^2+U^2)^{1/2}}=e^{2i\chi(\X,\lambda^2)}.
 \label{eq:polarization_direction}
\end{equation}
Angular statistics without Faraday rotation were explored by \citet{LazarianYuenPogosyan2022,LazarianPogosyanHu2026}. In the separated-screen geometry, \citet{Melnichenka2026} found with AthenaK simulations and synthetic fields that the polarization-direction spectrum responds strongly to fluctuations of the foreground rotation measure, particularly at moderate rotation. In contrast, internal Faraday rotation multiplies the emissivity at each depth by a different phase. The depth integral must be performed before the observed polarization is normalized. This distinction calls for a separate correlation theory and gives the present work its physical scope.

This difference makes the problem both physically interesting and statistically nontrivial. Ordinary polarization correlations weight the emission by polarized amplitude and depend on cancellation along the LOS. Normalizing each measured polarization vector removes that amplitude weighting, allowing changes in its phase to produce a pronounced directional response. The resulting direction correlation is nevertheless not the normalized ordinary correlation: the nonlinear operation $P/|P|$ must be averaged correctly. We derive the directional correlation by fixing the Faraday-depth histories of two sightlines before averaging over them. This separates the relative rotation of the net polarization from changes in the interference of emission at different depths. It also explains why a Gaussian distribution at each sky position alone is insufficient to predict directional correlations. We distinguish random Faraday decorrelation from coherent mean rotation and establish how the Faraday-density spectrum enters the observed statistics. The weak-rotation response has independent physical content because it contains intrinsic--Faraday cross terms that disappear under simplified independence assumptions.

The paper is organized as follows. Section~\ref{sec:formalism} defines the mixed-volume polarization and polarization-direction statistics. Section~\ref{sec:phase_mapping} develops the conditional theory, including a deep-volume analytical example. Section~\ref{sec:strong} derives the strong stochastic-rotation correlation and its spatial and spectral branches. Section~\ref{sec:mean} treats strong mean rotation and LOS Fourier selection, and Section~\ref{sec:interferometry} gives the interferometric interpretation. Section~\ref{sec:tangential} compares the polarization-direction spectrum with the normalized tangential gradient and presents its scaling regimes. Section~\ref{sec:lp16compare} corrects the LP16 strong-mean spatial and interferometric asymptotes and compares the LP16 polarization measures with the two direction-based measures. Section~\ref{sec:geometry} discusses how their combined use distinguishes emitting and rotating regions. Sections~\ref{sec:discussion} and \ref{sec:summary} discuss and summarize the results. Appendix~\ref{app:weak} develops the weak internal Faraday-rotation expansion. Appendix~\ref{app:math} provides step-by-step mathematical derivations of the main formulas.

\section{Polarization and its directional correlation}\label{sec:formalism}
Following LP16, define the intrinsic polarized emissivity $P_i(\X,z)=\epsilon(\X,z)e^{2i\psi(\X,z)}$, local Faraday-rotation density $\phi=\kappa n_e H_\parallel$, and accumulated depth $\Phi(\X,z)=\int_0^z\phi(\X,s)\,ds$. With $x\equiv\lambda^2$, the observed mixed-medium polarization is
\begin{equation}
 P(\X,x)=\int_0^L dz\,P_i(\X,z)e^{2ix\Phi(\X,z)}.
 \label{eq:Pmixed}
\end{equation}
We use the LP16 symbols $r_i,m$ for intrinsic-polarization correlations and $r_\phi,m_\phi,\bar\phi,\sigma_\phi$ for Faraday-density statistics. Smooth synchrotron-frequency dependence is factored out of $P_i$.

The direction-only correlation, structure function, and transverse spectrum are
\begin{align}
 \xi_u(\R,x)&=\avg{u(\X,x)u^*(\X+\R,x)},\nonumber\\
 D_u(\R,x)&=2[1-\Rea\,\xi_u(\R,x)],\label{eq:Du}\\
 P_u(\K,x)&=\avg{|\widetilde u(\K,x)|^2}.\label{eq:Pu}
\end{align}
The isotropic ring-averaged spectrum is written $P_u(K,x)$. The usual two-dimensional structure-function correspondence is $D_u\propto R^\beta\ \Longleftrightarrow\ P_u\propto K^{-(2+\beta)}$ for $0<\beta<2$.

\section{Directional correlations for prescribed Faraday depths}\label{sec:phase_mapping}
\subsection{Exact statistical relation for Gaussian emissivity}
Consider two sightlines at $\X_a$ ($a=1,2$) and fix their entire accumulated Faraday-depth histories $\Phi_a(z)$, which in general differ. Suppose that $P_i$ is a zero-mean circular complex Gaussian field, independent of $\phi$, with covariance $\xi_i^{ab}(z,z')=\avg{P_i(\X_a,z)P_i^*(\X_b,z')}$. Circularity here means that the joint distribution is invariant under a common rotation in the $Q$--$U$ plane, and $\avg{P_i P_i}=0$. Because Eq.~(\ref{eq:Pmixed}) is linear in $P_i$ for fixed $\Phi_a$, the two observed polarizations have an exactly Gaussian conditional distribution. Its covariance is
\begin{equation}
\begin{split}
 C_{ab}[\Phi]&=\int_0^L dz\int_0^L dz'\,\xi_i^{ab}(z,z')\\
 &\quad\times e^{2ix[\Phi_a(z)-\Phi_b(z')]} .
\end{split}\label{eq:conditional_cov}
\end{equation}
The corresponding normalized correlation is $\rho_c=C_{12}/\sqrt{C_{11}C_{22}}$, whenever both variances are positive. Write
\begin{equation}
 {\cal F}(\rho)=\frac{\pi}{4}\rho\,
 {}_2F_1\!\left(\frac12,\frac12;2;|\rho|^2\right).
 \label{eq:phase_gaussian_exact}
\end{equation}
Appendix~\ref{app:math:phase} derives this function by integrating the directions of a correlated complex Gaussian pair over its joint density. Its phase is that of $\rho$. The direction correlation for the mixed volume is then
\begin{equation}
 \xi_u(R,x)=\left\langle {\cal F}(\rho_c[\Phi])\right\rangle_{\Phi_1,\Phi_2} .
 \label{eq:conditional_main}
\end{equation}
This is exact for the specified emissivity statistics and does not require Gaussian Faraday-density fluctuations. The average is over two correlated depth histories, not merely their terminal rotation measures. A finite intrinsic emissivity correlation length is included directly in Eq.~(\ref{eq:conditional_cov}).

In contrast, ordinary polarization measures
\begin{equation}
 \rho_P=\frac{\avg{C_{12}}_\Phi}
 {\sqrt{\avg{C_{11}}_\Phi\avg{C_{22}}_\Phi}}.
 \label{eq:rho_unconditional}
\end{equation}
In general $\avg{{\cal F}(\rho_c)}_\Phi\ne{\cal F}(\rho_P)$. The function is nonlinear and the conditional variances can fluctuate. Even a Gaussian one-point distribution of $P$ does not establish Gaussianity of the pair. For example, let $P_2=e^{i\vartheta}P_1$ with circular Gaussian $P_1$ and independent random $\vartheta$. Both marginal distributions are Gaussian, but $\xi_u=\rho_P=\avg{e^{-i\vartheta}}$ instead of Eq.~(\ref{eq:phase_gaussian_exact}) applied to $\rho_P$. The latter is exact if the measured pair is itself jointly zero-mean circular Gaussian, as in pure deterministic mean rotation of a Gaussian source. It is also a useful approximation when the conditional covariances concentrate. At small conditional $|\rho_c|$, ${\cal F}(\rho_c)=(\pi/4)\rho_c+O(\rho_c|\rho_c|^2)$. Only if the conditional variances can be replaced by their mean values (or the relevant correlations with them are negligible) does this give
\begin{equation}
 \xi_u\simeq(\pi/4)\rho_P .\label{eq:phase_lowrho}
\end{equation}
Small $|\rho_P|$ alone is not sufficient.

\subsection{Short-correlated emissivity and neighboring sightlines}
An analytically tractable example is $\xi_i^{ab}(z,z')=A_i g_i(R_{ab})\delta(z-z')$, the limit of a longitudinal emissivity correlation length much shorter than the Faraday phase variation length. Let $g_i(0)=1$ and take $g_i(R)$ real and positive near $R=0$. Then $C_{aa}=A_i L$ and
\begin{equation}
 \rho_c=g_i(R)\frac{1}{L}\int_0^L dz\,e^{i\theta(z)},\quad
 \theta(z)=2x[\Phi_1(z)-\Phi_2(z)] .
 \label{eq:white_conditional}
\end{equation}
Equation~(\ref{eq:conditional_main}) with Eq.~(\ref{eq:white_conditional}) requires only realizations of the differential Faraday depth: both sightlines are evaluated at the same depth, so the common part $(\Phi_1+\Phi_2)/2$ of their Faraday depths cancels in $C_{12}$. It includes arbitrary rotation strength. This short-correlated emissivity example is distinct from the slowly varying emissivity used for the strong stochastic depth window in Section~\ref{sec:strong}.

For nearly coherent sightlines define $\bar\theta=L^{-1}\int dz\,\theta(z)$ and $V_\theta=L^{-1}\int dz\,[\theta(z)-\bar\theta]^2$. Expansion gives $\rho_c\simeq g_i e^{i\bar\theta}(1-V_\theta/2)$. Thus the relative rotation of the whole polarization is controlled by $\bar\theta$, while the change in interference between emitting depths is controlled by $V_\theta$. Put $a=1-g_i^2$. Since ${\cal F}(\rho_c)=e^{i\arg\rho_c}\,{\cal F}(|\rho_c|)$, Eq.~(\ref{eq:phase_gaussian_exact}) is needed only for the modulus. For $|\rho_c|\to1$ and $\varepsilon_c=1-|\rho_c|^2\to0$ it gives
\begin{equation}
 1-{\cal F}(|\rho_c|)=\frac {\varepsilon_c}{4}\left[\ln\frac{16}{\varepsilon_c}-1\right]
 +O({\varepsilon_c}^2\ln(1/\varepsilon_c)).\label{eq:Du_core_exact}
\end{equation}
At fixed Faraday phase $\arg\rho_c\simeq\bar\theta$ and $\varepsilon_c\simeq a+V_\theta$. Multiplying Eq.~(\ref{eq:Du_core_exact}) by $\cos\bar\theta$ for the real part and averaging over the Faraday phases gives
\begin{equation}
\begin{split}
 1-\Rea\,\xi_u\simeq&\ \tfrac12\avg{\bar\theta^2}\\
 &+\tfrac14\left\langle(a+V_\theta)
 \left[\ln\frac{16}{a+V_\theta}-1\right]\right\rangle .
\end{split}\label{eq:conditional_near}
\end{equation}
Here the phase differences and $a$ are small; the expansion holds in the ensemble when their scaled moments are finite. The logarithm multiplies the variations about the depth-averaged phase, not the entire differential rotation.

\subsection{Deep-volume limit}
If the differential Faraday density has a finite longitudinal integral scale and $L$ is much larger than that scale, its accumulated phase can approach a Brownian process. Set $\theta(Lt)=s W(t)$, $0\le t\le1$, where $W$ is standard Brownian motion and $s^2$ is the terminal phase variance. With $\bar W=\int_0^1 W(t)dt$ and $V=\int_0^1(W-\bar W)^2dt$, one has $\avg{\bar W^2}=1/3$ and $\avg V=1/6$ (Appendix~\ref{app:math:conditional}). Define $\varepsilon_F=s^2/2$, the Faraday contribution to $\varepsilon=1-\rho_P^2$ at leading order. Then $\varepsilon\simeq a+\varepsilon_F$, $\avg{\bar\theta^2}=2\varepsilon_F/3$, and $\avg{V_\theta}=\varepsilon_F/3$. At a fixed ratio of intrinsic and Faraday contributions,
\begin{equation}
 1-\Rea\,\xi_u\sim\frac14\left(a+\frac{\varepsilon_F}{3}\right)
 \ln\frac1\varepsilon .\label{eq:mixed_core}
\end{equation}
For $a=0$, the next term is calculable:
\begin{equation}
 1-\xi_u=\frac{\varepsilon}{12}
 \left[\ln\frac{16}{\varepsilon}+C_B\right]+o(\varepsilon).
 \label{eq:brownian_core}
\end{equation}
Here $C_B=3-\ln 2-6\avg{V\ln V}$, with an integral representation derived in Appendix~\ref{app:math:conditional}. The leading logarithmic coefficient is three times smaller than for a jointly Gaussian observed pair. Two thirds of the ordinary decorrelation is a relative rotation rather than a change in interference. Our starting model of independent short-correlated Gaussian emissivity makes the conditional calculation exact. For a deterministic uniform emitter the remaining randomness after fixing the differential phase is supplied by the common Faraday phase. Its Gaussian treatment relies on the central limit theorem and is an approximation for a volume much deeper than $r_\phi$. Appendix~\ref{app:math:conditional} derives the corresponding conditional correlation and its deep-volume limit.

The ratio of the logarithmic coefficients in Eq.~(\ref{eq:mixed_core}) and the Gaussian-pair prediction is $(a+\varepsilon_F/3)/(a+\varepsilon_F)$. Thus it lies between $1/3$ and unity in this model. The correction depends on which fluctuations control the measured spatial scale. For finite longitudinal emissivity correlations the two variances in Eq.~(\ref{eq:conditional_cov}) fluctuate, and neither this ratio nor the constant in Eq.~(\ref{eq:brownian_core}) is universal. The full conditional calculation retains these effects through the covariance of the emissivity. A regular plane-of-sky magnetic field can also produce a nonzero mean emissivity, and the physical synchrotron emissivity is generally not Gaussian. Such cases require the directional average with the appropriate conditional density, rather than Eq.~(\ref{eq:phase_gaussian_exact}).

\section{Strong turbulent Faraday rotation}\label{sec:strong}
\subsection{The coherence depth of ordinary polarization}\label{sec:stochasticwindow}
For Gaussian Faraday-density fluctuations independent of the emissivity, the LP16 ordinary polarization correlation is
\begin{equation}
\begin{split}
 \xi_P(R,x)&=\int_0^L dz_1\int_0^L dz_2\,\xi_i(R,\Delta z)\\
 &\quad\times e^{2ix\bar\phi\Delta z}
 e^{-4x^2D_{\Delta\Phi}(R,z_1,z_2)},
\end{split}\label{eq:LP16_CP_exact}
\end{equation}
where $\Delta z=z_1-z_2$, $z_+=(z_1+z_2)/2$, and $D_{\Delta\Phi}$ is half the variance of the difference of the fluctuating Faraday depths. Define the two characteristic depths
\begin{equation}
 \ell\equiv L_{\sigma_\phi}=(\sqrt2\,x\sigma_\phi)^{-1},\qquad
 L_{\bar\phi}=(x|\bar\phi|)^{-1}.
 \label{eq:Lfaraday}
\end{equation}
For $\ell\ll\min(r_i,r_\phi,L,L_{\bar\phi})$, the Gaussian integral over $\Delta z$ yields, away from the front and rear boundaries,
\begin{align}
 \xi_P(R,x)&\simeq\sqrt\pi\ell\,\xi_i(R,0)W_\phi(R),\label{eq:CP_strong_random}\\
 W_\phi(R)&\simeq\int_0^L dz_+\left[\frac{\Lambda_-}{2\sigma_\phi^2}\right]^{-1/2}
 e^{-4x^2D^+_{\Delta\Phi}}.\label{eq:Wphi_def}
\end{align}
Here $D^+_{\Delta\Phi}=D_{\Delta\Phi}(R,z_+,z_+)$, and $\Lambda_-/4$ is the quadratic coefficient in $\Delta z$. Its derivation, finite limits, and mean-rotation factor are given in Appendix~\ref{app:math:window}. $W_\phi(0)\simeq L$. When $W_\phi$ approaches $\ell$, the front boundary must be retained. The approximate factorization $\rho_P\simeq g_i(R)w_\phi(R)$ uses $w_\phi=W_\phi/L$, $g_i(R)=\xi_i(R,0)/\xi_i(0,0)$; neither that factorization nor $W_\phi$ alone determines $\xi_u$ in general.

\begin{figure*}
\centering
\includegraphics[width=0.94\textwidth]{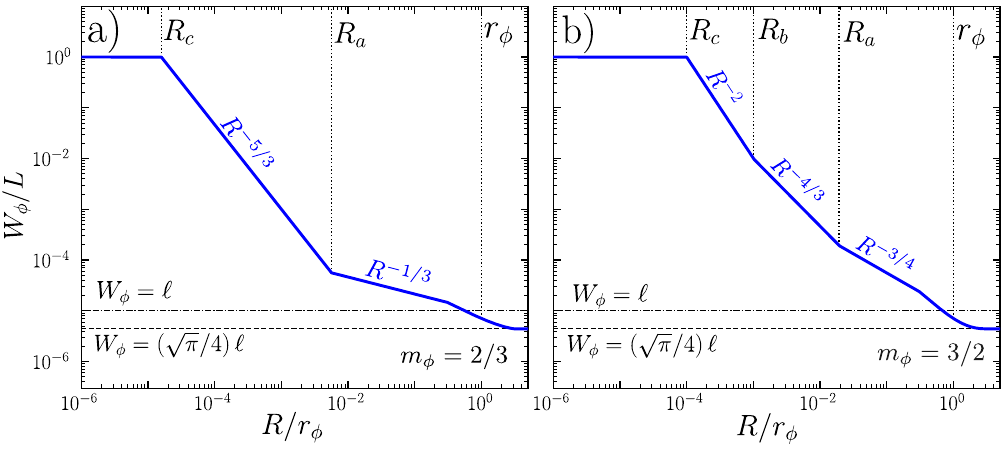}
\caption{Ordinary polarization coherence depth $W_\phi$ of a deep slab under strong stochastic Faraday rotation, built from the limiting branches of Eqs.~(\ref{eq:Wphi_branches}) and (\ref{eq:Wphi_steep}) with $\ell=10^{-3}r_\phi$ and $L=10^{2}r_\phi$. (a) Shallow Faraday-density spectrum, $m_\phi=2/3$: the branches $R^{-1-m_\phi}=R^{-5/3}$ and $R^{-m_\phi/2}=R^{-1/3}$. (b) Steep spectrum, $m_\phi=3/2$: the additional $R^{-2/m_\phi}=R^{-4/3}$ branch between $R_b\sim\ell$ and $R_a$ separates the $R^{-2}$ and $R^{-m_\phi/2}=R^{-3/4}$ intervals. Dotted lines mark $R_c$ and $R_a$ from Eqs.~(\ref{eq:Ra_strong})--(\ref{eq:Rc_strong}), $R_b$, and $r_\phi$. The dash-dotted line marks $W_\phi=\ell$: as $W_\phi$ approaches it, the front boundary must be retained. The dashed line marks the independent-sightline plateau $W_\phi=(\sqrt\pi/4)\,\ell$, the $c=0$ limit of Eq.~(\ref{eq:plateau}), onto which the curves are interpolated between $0.3\,r_\phi$ and $3\,r_\phi$. The curves join limiting scalings and are not evaluations of the finite-depth integral.}
\label{fig:strong_window}
\end{figure*}

For an isotropic Faraday-density covariance $C_\phi(s)\simeq\sigma_\phi^2[1-(s/r_\phi)^{m_\phi}]$, $0<m_\phi<2$, consider a deep volume $L\gg r_\phi$. The depth that contributes to the correlation is set by $4x^2D^+_{\Delta\Phi}\sim1$. For $0<m_\phi<1$, the principal depth ranges are
\begin{equation}
 W_\phi\sim\begin{cases}
 L & R\ll R_c,\\
 \ell^2 r_\phi^{m_\phi} R^{-1-m_\phi}&R_c\ll R\ll R_a,\\
 \ell(r_\phi/R)^{m_\phi/2}&R_a\ll R\ll r_\phi.
 \end{cases}\label{eq:Wphi_branches}
\end{equation}
For $1<m_\phi<2$ there is an additional interval, because the effective depth first becomes shorter than $r_\phi$ while remaining much larger than $R$:
\begin{equation}
 W_\phi\sim\begin{cases}
 L & R\ll R_c,\\
 \ell^2r_\phi/R^2 & R_c\ll R\ll R_b,\\
 r_\phi(\ell/R)^{2/m_\phi}&R_b\ll R\ll R_a,\\
 \ell(r_\phi/R)^{m_\phi/2}&R_a\ll R\ll r_\phi.
 \end{cases}\label{eq:Wphi_steep}
\end{equation}
Here $R_b\sim\ell$. Matching the depth to $R$ gives, for either range of $m_\phi$,
\begin{align}
 R_a&\sim r_\phi(\ell/r_\phi)^{2/(2+m_\phi)},\label{eq:Ra_strong}\\
 R_c&\sim\left(\ell^2r_\phi^{\widetilde m_\phi}/L\right)^{1/(1+\widetilde m_\phi)},\label{eq:Rc_strong}\\
 \widetilde m_\phi&=\min(m_\phi,1).\nonumber
\end{align}
The marginal $m_\phi=1$ has logarithmic corrections to these depth estimates. All matching coefficients here are of order unity. In particular the coefficient of the deep-volume branch approaches its small-$R$ limit slowly when $m_\phi$ is close to one. Appendix~\ref{app:math:window} gives a covariance-integral definition that avoids treating this coefficient as a constant at finite separations. The core radius is best determined from the finite-depth correlation rather than by matching unbounded power laws. For $R\gtrsim r_\phi$, the exact near-surface integral has a different normalization from the extended longitudinal Gaussian integral. For negligible mean rotation and $c=C_\phi(R)/\sigma_\phi^2$ bounded away from unity, the leading result is
\begin{equation}
 w_\phi(R)\simeq\frac{\ell}{L}
 \frac{\pi/2+\arcsin c}{2\sqrt\pi\sqrt{1-c^2}}.
 \label{eq:plateau}
\end{equation}
At independent sightlines, $w_\phi\simeq(\sqrt\pi/4)\ell/L$, half the result obtained by extending the longitudinal integral beyond the front face. For a uniform source and independent widely separated sightlines the plateau represents the squared mean polarization, not a nonzero-wavenumber contribution. For a fluctuating source it is further modulated by the transverse emissivity correlation. Figure~\ref{fig:strong_window} illustrates the corrected scale ordering.

\subsection{Expected directional spectra and finite scale ranges}\label{sec:branches}
If the low-coherence approximation in Eq.~(\ref{eq:phase_lowrho}) applies and $g_i(R)$ is nearly constant over the Faraday-dominated interval, the directional correlation inherits the above depth powers. For isotropic maps the spectrum is the Fourier transform of the connected correlation $\Xi_u$ (the mean contributes only at $K=0$):
\begin{equation}
 P_u(K)=2\pi\int_0^\infty R\,dR\,J_0(KR)\Xi_u(R).
 \label{eq:hankel_def}
\end{equation}
A correlation proportional to $R^{-\gamma}$ gives $P_u\propto K^{\gamma-2}$ when $1/2<\gamma<2$ and both boundaries of the interval are sufficiently distant. For $\gamma=2$, the corresponding leading term is logarithmic. The resulting conditional low-coherence predictions are listed in Table~\ref{tab:branches}. If the intrinsic factor varies as $R^{-a_i}$, add $a_i$ to $\gamma$ before examining convergence. At intermediate separations with $m_\phi\le1$ the finite range matters. A contribution from $R_a<R<r_\phi$ is proportional to
\begin{equation}
 K^{m_\phi/2-2}\int_{KR_a}^{Kr_\phi}dt\,t^{1-m_\phi/2}J_0(t),
 \label{eq:finite_window_hankel}
\end{equation}
which cannot in general be replaced by a constant times $K^{m_\phi/2-2}$.

The characteristic wavenumbers have the following physical meanings. The core wavenumber $K_c=1/R_c$ marks the transition from coherence over the entire emitting depth $L$ to coherence over only part of that depth. For $1<m_\phi<2$, $K_b=1/R_b\sim1/\ell$ marks the point at which the coherence depth becomes comparable to the Faraday-density correlation length $r_\phi$. The wavenumber $K_a=1/R_a$ marks the transition at which the coherence depth becomes comparable to the transverse separation itself. Finally, $K_\phi=1/r_\phi$ corresponds to the Faraday-density correlation length. Thus, for $1<m_\phi<2$ and well-separated scales, $K_\phi\ll K_a\ll K_b\ll K_c$. For $0<m_\phi<1$, there is no separate $K_b$ break.

\begin{deluxetable*}{llll}
\tabletypesize{\normalsize}
\tablecaption{Strong stochastic rotation: limiting depth and directional spectrum\label{tab:branches}}
\tablewidth{0pt}
\tablehead{\colhead{Range in $R$}&\colhead{Range in $K$}&\colhead{$W_\phi(R)$}&\colhead{$P_u(K)$}}
\startdata
$R_c\ll R\ll R_a$, $m_\phi<1$ & $K_a\ll K\ll K_c$ & $R^{-1-m_\phi}$ & $K^{-1+m_\phi}$ \\
$R_c\ll R\ll R_b$, $m_\phi>1$ & $K_b\ll K\ll K_c$ & $R^{-2}$ & $\ln(K_c/K)+\mathrm{const}$ \\
$R_b\ll R\ll R_a$, $m_\phi>1$ & $K_a\ll K\ll K_b$ & $R^{-2/m_\phi}$ & $K^{-2+2/m_\phi}$ \\
$R_a\ll R\ll r_\phi$, $m_\phi>1$ & $K_\phi\ll K\ll K_a$ & $R^{-m_\phi/2}$ & $K^{-2+m_\phi/2}$ \\
$R_a\ll R\ll r_\phi$, $m_\phi<1$ & $K_\phi\ll K\ll K_a$ & $R^{-m_\phi/2}$ & Eq.~(\ref{eq:finite_window_hankel}) \\
\enddata
\tablecomments{The entries give scale dependences for $0<m_\phi<2$, excluding the marginal case $m_\phi=1$. The spectral powers require low coherence, slowly varying intrinsic correlations, and well-separated transition scales. The last row requires the full finite-range spectrum rather than a universal power law.}
\end{deluxetable*}

The breaks move with wavelength as
\begin{equation}
 \begin{split}
 K_a&\propto\lambda^{4/(2+m_\phi)},\quad K_c\propto\lambda^{4/(1+\widetilde m_\phi)},\\
 K_b&\propto\lambda^2\quad (1<m_\phi<2).
 \end{split}\label{eq:break_lambda}
\end{equation}
These wavelength dependences neglect logarithmic corrections at $m_\phi=1$. For example, at $m_\phi=2/3$ the inner slope is $-1/3$ for $P_u$, and at $m_\phi=3/2$ the new branch would give $-2/3$ for $P_u$ if the required statistical and scale orderings hold. These are limiting slopes. The two scale boundaries can affect almost the entire nominal power-law interval. For $m_\phi>1$, the logarithmic inner form also retains a finite additive term. The accuracy of the low-coherence directional approximation and the required scale separation depend on the statistics of the emitting and rotating fields. A full prediction for Gaussian emissivity is obtained from Eq.~(\ref{eq:conditional_main}) followed by Eq.~(\ref{eq:hankel_def}), rather than by applying ${\cal F}$ to the averaged ordinary coherence. The core gradient predictions are discussed in Section~\ref{sec:tangential}.

\section{Strong mean rotation and Fourier selection}\label{sec:mean}
When the regular Faraday rotation dominates, the phase $e^{2ix\bar\phi\Delta z}$ oscillates but does not suppress distant separations in absolute value. The replacement $\xi_P\sim LL_{\bar\phi}\xi_i(R,0)$ in LP16 Eq.~(57) is therefore not a controlled stationary-phase approximation. Section~\ref{sec:lp16correction} derives the corrected expression from the finite-slab integral and gives the corresponding correction to LP16's interferometric spectrum.

For negligible fluctuating Faraday density, define $k_\parallel=2x\bar\phi$. The \emph{exact} finite-slab source correlation is
\begin{equation}
 \xi_P(R,x)=\int_{-L}^L ds\,(L-|s|)e^{ik_\parallel s}\xi_i(R,s).
 \label{eq:CP_strong_mean}
\end{equation}
When the source fluctuations dominate the contribution from the front and back boundaries of the emitting slab,
\begin{equation}
 \xi_P(R,x)\simeq L\widehat\xi_i(R,k_\parallel),\quad
 \widehat\xi_i=\int_{-\infty}^\infty ds\,e^{ik_\parallel s}\xi_i(R,s).
 \label{eq:CP_mean_Fourier}
\end{equation}
For $\xi_i(0,0)-\xi_i(0,s)\simeq C_i|s|^m$ with $0<m<1$, this gives
\begin{equation}
 \avg{|P|^2}\simeq 2LC_i\Gamma(1+m)
 \sin\!\left(\frac{\pi m}{2}\right)|k_\parallel|^{-1-m}
 \propto\lambda^{-2(1+m)}.
 \label{eq:mean_variance}
\end{equation}
For $0<m<1$, this small-separation behavior of the source correlation dominates the $k_\parallel^{-2}$ contribution from the sharp boundaries of the slab. For smoother source correlations or different depth profiles, the original finite-slab integral should be used.

\begin{figure*}
\centering
\includegraphics[width=0.88\textwidth]{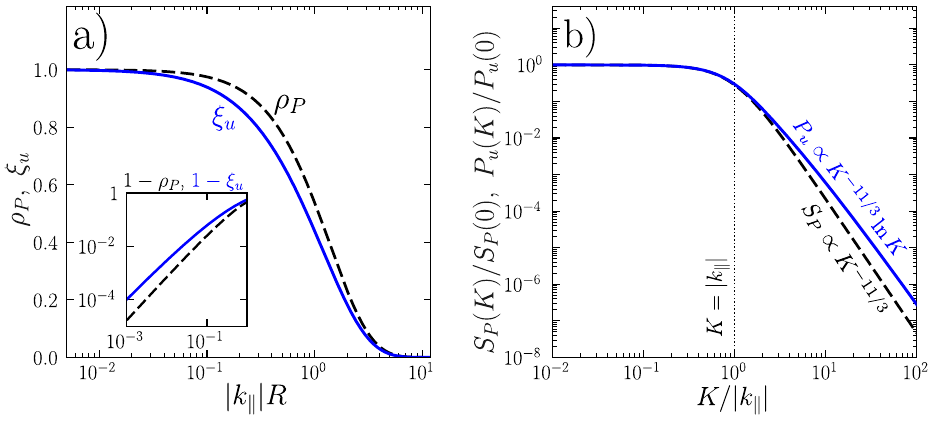}
\caption{Pure mean rotation for the isotropic source spectrum with $m=2/3$. Black dashed curves show ordinary polarization and blue solid curves the polarization direction. (a) Ordinary coherence $\rho_P$, Eq.~(\ref{eq:rho_strong_mean}), and directional correlation $\xi_u={\cal F}(\rho_P)$, Eq.~(\ref{eq:phase_gaussian_exact}), against $|k_\parallel|R$. The inset shows $1-\rho_P\propto(|k_\parallel|R)^{1+m}$ and $1-\xi_u$, which carries the additional logarithm of Eq.~(\ref{eq:mean_direction}), on logarithmic axes. (b) Ordinary interferometric spectrum $S_P(K)\propto(K^2+k_\parallel^2)^{-(3+m)/2}$ and direction spectrum $P_u(K)$, evaluated numerically from Eq.~(\ref{eq:hankel_def}) with $\xi_u$, each normalized by its value at $K=0$. At high $K$, $S_P\propto K^{-(3+m)}=K^{-11/3}$ and $P_u\propto K^{-11/3}\ln K$. The dotted line marks $K=|k_\parallel|=2\lambda^2|\bar\phi|$.}
\label{fig:strong_realspace}
\end{figure*}

For an isotropic three-dimensional source spectrum proportional to $(K^2+k_\parallel^2)^{-(3+m)/2}$, the normalized transverse correlation is
\begin{align}
 \rho_P(R)&=\frac{2^{1-\nu}}{\Gamma(\nu)}y^\nu K_\nu(y),\label{eq:rho_strong_mean}\\
 y&=|k_\parallel|R,\qquad \nu=(1+m)/2,\nonumber
\end{align}
where $K_\nu$ is a modified Bessel function. It decreases on the scale $R\sim|k_\parallel|^{-1}\propto\lambda^{-2}$. For the independent circular Gaussian emissivity model, the Faraday phase is now deterministic and Eq.~(\ref{eq:conditional_main}) reduces exactly to $\xi_u={\cal F}(\rho_P)$. Its leading small-$R$ behavior for $0<m<1$ becomes
\begin{equation}
 D_u\propto(R/L_{\bar\phi})^{1+m}\ln(L_{\bar\phi}/R),
 \quad P_u\propto K^{-3-m}\ln(KL_{\bar\phi}).
 \label{eq:mean_direction}
\end{equation}
The logarithm and the nonanalytic source term determine the leading high-$K$ behavior. Analytic terms in $1-\rho_P$, although subleading at small $R$, generate further nonanalytic contributions after applying ${\cal F}$. Their spectral corrections can converge slowly. For quantitative fits at finite $K$, use Eqs.~(\ref{eq:rho_strong_mean}), (\ref{eq:phase_gaussian_exact}), and (\ref{eq:hankel_def}), rather than the leading power alone. Figure~\ref{fig:strong_realspace} shows $\rho_P$, $\xi_u={\cal F}(\rho_P)$, and the resulting spectra: at high $K$ the direction spectrum $P_u$ exceeds $S_P\propto K^{-(3+m)}$ by a slowly growing logarithmic factor. Appendix~\ref{app:math:mean} explains this behavior and provides a finite-beam calculation. These particular transverse exponents assume an isotropic source. In MHD turbulence the selected $k_\parallel$ instead probes the \emph{anisotropic} three-dimensional polarized-emissivity spectrum.

\section{Interferometric implications}\label{sec:interferometry}
The spectrum $S_P(\K,x)$ of the ordinary complex polarization differs from the spectrum $P_u(\K,x)$ of its normalized direction. For pure mean rotation, the spectrum observed by an interferometer averages the intrinsic source spectrum over a narrow range of line-of-sight wavenumbers set by the emitting slab depth. When its width is sufficiently narrow,
\begin{equation}
 S_P(\K,x)\simeq L S_i(\K,k_\parallel),
 \qquad k_\parallel=2x\bar\phi.
 \label{eq:SP_selection}
\end{equation}
For an isotropic inertial-range source, $S_P(K,x)\propto L(K^2+k_\parallel^2)^{-(3+m)/2}$. The transverse break therefore moves as $K_{\rm br}\propto\lambda^2$, with low-$K$ flattening and a $K^{-(3+m)}$ high-$K$ slope. For anisotropic turbulence the wavelength-dependent shape of $S_i(\K,k_\parallel)$ contains additional information about the magnetic-field geometry. The exact finite-slab convolution, necessary near boundaries and for narrow source features, is derived in Section~\ref{sec:lp16correction}.

\section{Comparison with normalized polarization gradient}\label{sec:tangential}
The spatial spectrum of polarization directions and the normalized tangential polarization gradient are two ways to analyze the same amplitude-normalized Stokes maps. The first measures how polarization directions remain correlated over a range of spatial scales. The second measures their local spatial variation. Both respond to Faraday rotation without giving additional weight to intrinsically bright polarized regions, and both can be obtained from a single-wavelength pair of Stokes maps. Their different weighting of spatial scales determines which is more convenient for an observation.

The direction spectrum $P_u(K)$ retains the relative contributions of large and small scales and makes changes of spectral slope or wavelength-dependent break scales easier to identify. It also avoids taking a spatial derivative of a noisy map. Its limitation is that weak small-scale Faraday fluctuations may be less prominent against power from larger-scale intrinsic polarization structure. The normalized tangential gradient $\mathcal G_u$ emphasizes rapid local changes in the polarization angle and is useful for locating small-scale Faraday structure. Its vector-gradient spectrum gives two additional powers of $K$ to the same fluctuations. This emphasis also makes the gradient more sensitive to beam smoothing, pixel noise, and the procedure used to normalize weak polarized signals. The spectrum of the \emph{vector} gradient contains no independent two-point information beyond $P_u(K)$; the spectrum of the \emph{scalar magnitude} $|\nabla u|$ is a different, nonlinear statistic.

For quantitative work, the direction spectrum is generally preferable for recovering broad spectral ranges and identifying the rotation-dependent break scales. The gradient can provide a complementary way to display small-scale Faraday variations when the angular resolution and polarized signal-to-noise are sufficient. The following relations connect the two techniques and collect their weak-, strong-turbulent-, and strong-mean-rotation predictions in Table~\ref{tab:tangential_regimes}.

Table~\ref{tab:tangential_regimes} summarizes the limiting gradient spectra. As in Table~\ref{tab:branches}, the strong stochastic rows describe sufficiently broad spatial intervals, not universal power laws at their transition scales.

\begin{deluxetable*}{lll}
\tabletypesize{\normalsize}
\tablecaption{Polarization-direction gradient: limiting spectral regimes\label{tab:tangential_regimes}}
\tablewidth{0pt}
\tablehead{\colhead{Faraday regime} & \colhead{Wavenumber interval} & \colhead{$P_{\nabla u}(K)$ scaling}}
\startdata
Weak, uniform emitter & $\Theta_F(1/K)\ll1$, $0<m_\phi<2$ & $\lambda^4 K^{-(1+m_\phi)}$ \\
Turbulent, coherent core ($m_\phi<1$) & $K\gg K_c$ & $K^{-(1+m_\phi)}[\ln(KR_c)+c_\beta]$ \\
Turbulent, coherent core ($m_\phi=1$) & $K\gg K_c$ & $K^{-2}\ln(Kr_\phi)$ \\
Turbulent, coherent core ($m_\phi>1$) & $K\gg K_c$ & $K^{-2}$ \\
Turbulent, low coherence ($m_\phi<1$) & $K_a\ll K\ll K_c$ & $K^{1+m_\phi}$ \\
Turbulent, low coherence ($m_\phi>1$) & $K_b\ll K\ll K_c$ & $K^2[\ln(K_c/K)+c]$ \\
Turbulent, low coherence ($m_\phi>1$) & $K_a\ll K\ll K_b$ & $K^{2/m_\phi}$ \\
Turbulent, low coherence ($m_\phi>1$) & $K_\phi\ll K\ll K_a$ & $K^{m_\phi/2}$ \\
Pure mean rotation & $KL_{\bar\phi}\gg1$, $0<m<1$ & $K^{-(1+m)}[\ln(KL_{\bar\phi})+c_0+\cdots]$ \\
\enddata
\tablecomments{The break wavenumbers $K_a$, $K_b$, $K_c$ and $K_\phi$ are defined in Section~\ref{sec:branches}. The weak-rotation row gives the Faraday contribution for a thick slab with a coherent intrinsic direction. $\Theta_F$ is defined in Appendix~\ref{app:weak}. Strong turbulent low-coherence rows require the conditions of Table~\ref{tab:branches}. The constant $c$ accounts for finite terms in the logarithmic branch. Coherent-core rows use the Faraday-dominated Brownian limit for short-correlated Gaussian emissivity. The mean-rotation row assumes an isotropic Gaussian source; its constant is $c_0=c_*-\ln2$, with $c_*$ given in Eq.~(\ref{eq:math:mean_correction}). For $D_u\propto R^\beta\ln(R_c/R)$, $c_\beta=-A'(\beta)/A(\beta)$ with $A$ of Eq.~(\ref{eq:math:powertransform}). For $m_\phi\le1$, the interval $K_\phi\ll K\ll K_a$ requires the finite-range transform.}
\end{deluxetable*}

\subsection{Normalized tangential gradient and its spectral regimes}
The radial--tangential decomposition of the polarization directional derivative \citep{Herron2018a,Herron2018b} is particularly simple when normalized by the polarized amplitude. The maximum tangential component of the spatial derivative of the Stokes vector is
\begin{equation}
 \left(\frac{\partial\boldsymbol P}{\partial s}\right)_{\!\mathrm{tang,max}}
 =2|P|\,|\nabla\chi|.
 \label{eq:tangential_herron}
\end{equation}
Here $\boldsymbol P=(Q,U)$ and $s$ denotes a direction on the sky. For $u=P/|P|=e^{2i\chi}$, differentiation gives
\begin{equation}
 \nabla u=2iu\nabla\chi,\qquad
 \mathcal G_u\equiv|\nabla u|=2|\nabla\chi|.
 \label{eq:tangential_def}
\end{equation}
Thus $\mathcal G_u$ measures how rapidly the polarization direction changes across the sky, after the polarized intensity has been divided out. Its relation to the polarization-direction correlation is exact.

For the \emph{vector} gradient, Fourier transformation of $\nabla u$ gives the exact relation
\begin{equation}
 P_{\nabla u}(K,\lambda^2)=K^2P_u(K,\lambda^2).
 \label{eq:tangential_spectrum}
\end{equation}
This statement concerns the spectrum of $\nabla u$, not the spectrum of the nonlinear scalar field $|\nabla u|$; the distinction matters when making numerical or observational comparisons.

With the weak-rotation phase $\chi=\chi_0+\lambda^2q+\lambda^4s+O(\lambda^6)$ from Appendix~\ref{app:weak},
\begin{align}
 \mathcal G_u^2={}&4|\nabla\chi_0|^2
  +8\lambda^2\nabla\chi_0\cdot\nabla q\nonumber\\
 &+4\lambda^4\left[|\nabla q|^2
       +2\nabla\chi_0\cdot\nabla s\right]+O(\lambda^6).
 \label{eq:tangential_weak}
\end{align}
The $\lambda^2$ term tests a source--Faraday cross correlation. For a uniform coherent emitter in a sufficiently thick slab, the isolated weak Faraday term is
\begin{equation}
 P_{\nabla u,F}=4\lambda^4 K^2P_q(K)
 \propto\lambda^4 K^{-(1+m_\phi)},\quad 0<m_\phi<2.
 \label{eq:tangential_weak_slope}
\end{equation}
Appendix~\ref{app:weak} derives the spectrum directly from the projected Faraday density; the analytic $R^2$ term of a steep-spectrum structure function does not give a $K^{-2}$ gradient spectrum for $m_\phi>1$. The vector-gradient identity is exact. The weak-phase expansion is a local Taylor expansion, and its averaged coefficients require the corresponding moments to exist.

In strong stochastic rotation, the small-separation $D_u$ must in general be calculated from Eq.~(\ref{eq:conditional_main}). For the short-correlated Gaussian emissivity model its near-coherence form is Eq.~(\ref{eq:conditional_near}), which includes both relative rotation and changes in interference. If $D_u\propto R^\beta\ln(R_c/R)$ with $0<\beta<2$, then the leading large-$K$ spectrum and beam-smoothed gradient variance scale as
\begin{align}
 P_{\nabla u}&\propto K^{-\beta}[\ln(KR_c)+c_\beta],\nonumber\\
 \avg{|\nabla u_b|^2}&\propto b^{\beta-2}\ln(R_c/b).
 \label{eq:tangential_core}
\end{align}
Here $b$ denotes the rms width of a Gaussian beam applied to the normalized polarization field. It is not a beam applied to $Q$ and $U$ before normalization. The amplitude must be derived from the appropriate directional correlation, not from the ordinary coherence alone. For example, in the Faraday-dominated Brownian limit, put $\varepsilon=1-\rho_P^2\simeq(R/R_0)^\beta$, with $R_0$ defined by this small-separation coefficient, not by a matching of power laws. Equation~(\ref{eq:brownian_core}) then gives
\begin{equation}
 D_u\simeq\frac{(R/R_0)^\beta}{6}
 \left[\beta\ln(R_0/R)+\ln16+C_B\right].
 \label{eq:core_brownian_spatial}
\end{equation}
For $0<\beta<2$ the leading logarithmic spectral amplitude is one third of the unconditional Gaussian-pair prediction at the same $\varepsilon$. At finite coherence the constant in Eq.~(\ref{eq:core_brownian_spatial}) matters. For $\beta=2$ a single real-space logarithm yields $P_{\nabla u}\propto K^{-2}$, with a logarithmic beam variance, not $K^{-2}\ln K$. For $m_\phi=1$ in the Faraday-dominated Brownian model, $\varepsilon\propto R^2\ln(r_\phi/R)$ instead. The leading $D_u$ is proportional to $R^2\ln^2(r_\phi/R)$, giving $P_{\nabla u}\propto K^{-2}\ln(Kr_\phi)$ and a squared logarithm in the beam variance. The coefficients for a long-correlated or non-Gaussian emitting field must be evaluated with its own directional statistics. These Fourier statements are derived in Appendix~\ref{app:math:hankel}. The low-coherence strong stochastic branches of Table~\ref{tab:branches} gain two powers of $K$; in particular $K^{1+m_\phi}$ for the inner $m_\phi<1$ interval and $K^{2/m_\phi}$ for the additional steep-spectrum interval. All retain the statistical and scale-separation requirements of that table.

For pure mean rotation of an isotropic circular Gaussian source with $0<m<1$, Eq.~(\ref{eq:mean_direction}) gives $P_{\nabla u}\propto K^{-(1+m)}\ln(KL_{\bar\phi})$ asymptotically. The finite-$K$ spectrum is obtained by transforming ${\cal F}(\rho_P)$ with Eq.~(\ref{eq:rho_strong_mean}) and multiplying by $K^2$. In practice the limiting logarithm converges slowly; Appendix~\ref{app:math:mean} gives the first correction. For a Gaussian beam the exact prediction in this model is
\begin{equation}
 \avg{|\nabla u_b|^2}=\frac{1}{2\pi}\int_0^\infty dK\,K^3P_u(K)e^{-b^2K^2}.
 \label{eq:beam_exact}
\end{equation}
The limiting dependence for $0<m<1$ is $L_{\bar\phi}^{-1-m}b^{m-1}\ln(L_{\bar\phi}/b)$, with slow convergence of both amplitude and slope. The first constant and a direct finite-$b$ evaluation are given in Appendix~\ref{app:math:mean}. For anisotropic sources use the Fourier-selected source spectrum and then the direction mapping. Adequate small-scale resolution is especially important for the gradient statistic.

\section{Comparison with measures in LP16}\label{sec:lp16compare}
The complex polarization and its wavelength derivative introduced in LP16 and the direction-based statistics developed here are complementary ways to probe the same emitting and rotating volume. We first correct the LP16 strong-mean spatial and interferometric asymptotes. We then compare the measures under weak, strong turbulent, and strong mean Faraday rotation, using the weak-rotation expansion in Appendix~\ref{app:weak}. The LP16 measures retain the polarization amplitude, whereas $u=P/|P|$ describes only its direction. The normalized tangential gradient, $\mathcal G_u=|\nabla u|$, gives additional weight to small-scale changes of that direction.

\subsection{Corrected LP16 strong-mean spatial and interferometric asymptotes}\label{sec:lp16correction}
\subsubsection{Why the mean-rotation cosine is not a depth window}\label{sec:lp16correction_cos}
For a uniform Faraday density, $P(\X,x)=\int_0^L P_i(\X,z)e^{ik_\parallel z}\,dz$ with $k_\parallel=2x\bar\phi$. Equation~(\ref{eq:CP_strong_mean}) follows exactly by combining the two longitudinal integrations. The phase has unit modulus, has no stationary point in $s$, and cannot be replaced by a positive window of width $L_{\bar\phi}$.

A constant source correlation $\xi_i(R,s)=C(R)$ makes the cancellation explicit:
\begin{equation}
 \int_{-L}^{L}ds\,(L-|s|)C(R)\cos(k_\parallel s)
 =\frac{2C(R)}{k_\parallel^2}[1-\cos(k_\parallel L)].
 \label{eq:cosexample}
\end{equation}
Keeping only the first cosine lobe would incorrectly give order $LC(R)/|k_\parallel|$. Restricting $R>L_{\bar\phi}$ does not fix the approximation. The surviving contribution depends on the longitudinal Fourier structure of the source and, in a sharp finite slab, on its faces.

For $R=0$, let $\xi_i(0,0)-\xi_i(0,s)=C_i|s|^m+o(|s|^m)$ with $0<m<2$. The nonanalytic source term yields the oscillatory transform
\begin{equation}
 \int_{-\infty}^\infty |s|^m e^{iks}\,ds
 =-2\Gamma(1+m)\sin\!\left(\frac{\pi m}{2}\right)|k|^{-(1+m)}
 \label{eq:cusptransform}
\end{equation}
(in the regularized Fourier sense), so the two minus signs give the positive variance in Eq.~(\ref{eq:mean_variance}). A sharp slab also contributes terms of order $k_\parallel^{-2}$, for example $2[\xi_i(R,0)-\cos(k_\parallel L)\xi_i(R,L)]/k_\parallel^2$ for a real even source. Thus $0<m<1$ is the clean hard-slab source-dominated regime. This both corrects the LP16 spatial approximation and recovers the strong-mean polarization-frequency exponent already obtained in its wavelength analysis.

\subsubsection{Finite-depth interferometric spectrum}\label{sec:lp16correction_interferometry}
Define the intrinsic three-dimensional spectrum by
\begin{equation}
 \xi_i(\R,s)=\int\frac{d^2\K}{(2\pi)^2}\frac{dq}{2\pi}\,
 S_i(\K,q)e^{i\K\cdot\R-iqs}.
 \label{eq:source_spec}
\end{equation}
Fourier transformation of the exact finite-slab correlation gives
\begin{align}
 S_P(\K,x)&=\int\frac{dq}{2\pi}S_i(\K,q)
 |W_L(k_\parallel-q)|^2,\label{eq:SPexact}\\
 W_L(p)&=Le^{ipL/2}\operatorname{sinc}(pL/2).
 \label{eq:window}
\end{align}
The squared slab Fourier window integrates to $2\pi L$. For a source spectrum that varies little over its width, it selects a narrow range of line-of-sight wavenumbers. This gives Eq.~(\ref{eq:SP_selection}), replacing the fixed strong-mean slopes quoted in LP16 Eqs.~(103)--(104). For an isotropic power-law source,
\begin{equation}
 S_P(K,x)\simeq LA(K^2+k_\parallel^2)^{-(3+m)/2}.
 \label{eq:SPisotropic}
\end{equation}
Its break lies at $K\sim|k_\parallel|\propto\lambda^2$; the low-$K$ branch flattens and the high-$K$ branch follows $K^{-(3+m)}$, provided the inertial range includes both. Appendix~\ref{app:math:mean} derives the Fourier window, transverse Bessel correlation, and variance scaling. Integrating over transverse wavenumber reproduces Eq.~(\ref{eq:mean_variance}). The result for ordinary polarization $S_P$ does not require a Gaussian probability distribution for the measured polarization. For deterministic mean rotation of a circular Gaussian emissivity field, the conversion to $P_u$ by Eq.~(\ref{eq:phase_gaussian_exact}) is exact. Other emissivity distributions need their own directional averages.

\subsubsection{Magnetic anisotropy}
Equation~(\ref{eq:SP_selection}) suggests sampling $S_i(\K,q)$ along its longitudinal wavenumber by changing wavelength. The isotropic $K_{\rm br}\propto\lambda^2$ example is not universal. For illustration, if the LOS is parallel to the mean magnetic field and $|k_\parallel|\sim L_0^{-1/3}K_\perp^{2/3}$, then
\begin{equation}
 K_{\rm br}\sim L_0^{1/2}|k_\parallel|^{3/2}\propto\lambda^3.
 \label{eq:cbexample}
\end{equation}
Finite-window effects, viewing angle, and the intrinsic polarized-emissivity tensor must be included in quantitative fits. This is a spectral selection in three-dimensional wavenumber, not a direct measurement of physical distance along the LOS.

\subsection{Comparison of the spatial, frequency, and derivative measures}\label{sec:lp16measures}
LP16 use the ordinary polarization spatial correlation $C_P(R,x)=\avg{P(\X,x)P^*(\X+\R,x)}$, its transverse spectrum $S_P(K,x)$, and the variation of these quantities with $x=\lambda^2$. A related measure is the correlation of $P_x\equiv\partial P/\partial x$. Their wavelength dependence is not interchangeable with the transverse spectral dependence of $P_u(K,x)$.

Using the weak-rotation expansion in Appendix~\ref{app:weak}, we can connect these observables directly. Denote by $M_0=P_0$ and $M_1,M_2$ the depth moments in Eq.~(\ref{eq:weakP}), and write $C_{ab}(R)=\avg{M_a(\X)M_b^*(\X+\R)}$. Multiplying the two Taylor series gives
\begin{equation}
\begin{split}
C_P(R,x)={}&C_{00}+2ix(C_{10}-C_{01})\\
&+x^2(4C_{11}-2C_{20}-2C_{02})+O(x^3).
\end{split}
\label{eq:LP16weakCP}
\end{equation}
By contrast, differentiation before taking the correlation yields
\begin{equation}
 P_x=2iM_1-4xM_2+O(x^2),\qquad
 C_{P_x}(R,0)=4C_{11}(R).
 \label{eq:LP16weakderivative}
\end{equation}
Thus the derivative measure retains a Faraday-weighted signal even as $x\to0$, provided the rotating medium is kept fixed. This is distinct from setting $\phi=0$, which also removes that signal. Neither Eq.~(\ref{eq:LP16weakCP}) nor Eq.~(\ref{eq:LP16weakderivative}) requires Gaussian Faraday statistics.

For a uniform emitter with one intrinsic polarization direction, $M_1=P_0q$ with $q$ given by Eq.~(\ref{eq:qweight}). At nonzero transverse wavenumber the isolated Faraday parts satisfy
\begin{equation}
\begin{split}
 S_{P_x}(K,0)&=4|P_0|^2P_q(K),\\
 P_{u,F}(K,x)&=\frac{x^2}{|P_0|^2}S_{P_x}(K,0),\\
 P_{\nabla u,F}(K,x)&=\frac{x^2K^2}{|P_0|^2}S_{P_x}(K,0).
\end{split}
\label{eq:LP16weaklink}
\end{equation}
All three therefore measure the same projected Faraday-density exponent for this simple source. The direction and gradient measures carry a factor $x^2=\lambda^4$, whereas the LP16 derivative has a finite short-wavelength limit. In a general emitting volume, the cross terms of Eq.~(\ref{eq:Duweak_exact}) prevent this simple identification and the full weighted statistics must be retained.

Table~\ref{tab:LP16spatial} summarizes the ordinary polarization spatial regimes. In strong turbulent rotation $W_\phi(R)$ describes the ordinary polarization coherence, not in general the normalized direction correlation; the latter requires Eq.~(\ref{eq:conditional_main}). Table~\ref{tab:LP16frequency} collects the LP16 wavelength and derivative measures, with the assumptions controlling their limiting behavior.

\begin{deluxetable*}{lll}
\tabletypesize{\normalsize}
\tablecaption{LP16 ordinary polarization: spatial regimes\label{tab:LP16spatial}}
\tablewidth{0pt}
\tablehead{\colhead{Regime} & \colhead{Complex-polarization statistic} & \colhead{Information and conditions}}
\startdata
Weak internal rotation & $C_P=C_{00}+O(x)$ & Intrinsic source dominates; see Eq.~(\ref{eq:LP16weakCP}) \\
Strong turbulent rotation & $C_P\simeq\sqrt\pi\,\ell\,\xi_i(R,0)W_\phi(R)$ & Emissivity correlation and Faraday coherence depth \\
Pure strong mean rotation & $S_P(\K,x)\simeq L S_i(\K,k_\parallel)$ & Source spectrum at $k_\parallel=2x\bar\phi$ \\
Separate foreground screen & $C_P=C_i\,\avg{e^{2ix\Delta\Phi_f}}$ & Factorization if source and screen are independent \\
\enddata
\tablecomments{The strong-turbulent expression assumes a sufficiently deep source, a short longitudinal Faraday window, and the LP16 Gaussian rotation model. Strong mean rotation requires finite-slab boundary terms to be subdominant. $x=\lambda^2$ and $C_i$ is the intrinsic polarization correlation. The complex-polarization spectrum is amplitude weighted, unlike $P_u$ and $P_{\nabla u}$.}
\end{deluxetable*}

\begin{deluxetable*}{lll}
\tabletypesize{\normalsize}
\tablecaption{LP16 wavelength and derivative statistics\label{tab:LP16frequency}}
\tablewidth{0pt}
\tablehead{\colhead{Measure} & \colhead{Limiting result} & \colhead{Principal requirement or use}}
\startdata
$C_{P_x}(R,0)$, weak rotation & $4\avg{M_1(\X)M_1^*(\X+\R)}$ & Multi-frequency derivative; Faraday-weighted source \\
$S_{P_x}(K,0)$, uniform emitter & $4|P_0|^2P_q(K)$ & $K^{-3-m_\phi}$ for an isotropic thick slab \\
$\avg{|P|^2}$, strong turbulent rotation & $\propto L\ell\propto\lambda^{-2}$ & Gaussian rotation, source smooth over $\ell$ \\
$\avg{|P|^2}$, pure strong mean rotation & $\propto\lambda^{-2(1+m)}$ & $0<m<1$; source-dominated Fourier term \\
$P_x$, arbitrary rotation & $2i\int dz\,\Phi P_i e^{2ix\Phi}$ & Faraday-depth weighting; no universal slope \\
\enddata
\tablecomments{The two variance rows are polarization-frequency asymptotes, not transverse power laws. At strong rotation the derivative correlation retains the full depth-weighted phase integral and is generally not obtained by differentiating a single-wavelength spatial spectrum. $P_{\nabla u}(K,x)=K^2P_u(K,x)$ follows exactly from Section~\ref{sec:tangential}; no frequency differentiation is needed for either normalized measure.}
\end{deluxetable*}

The contrast is particularly clear at weak rotation. The LP16 derivative directly weights each emitting contribution by its accumulated Faraday depth and remains sensitive when $\lambda^2$ is small, although it requires sufficiently accurate frequency sampling to estimate $P_x$. The direction spectrum can be measured from a single Stokes map and emphasizes rotation of the net polarization rather than its brightness, but its isolated Faraday term is suppressed by $\lambda^4$. The normalized gradient further favors small-scale structure through the factor $K^2$, at the cost of greater sensitivity to angular resolution, beam smoothing, and noise. At the level of two-point statistics it contains no information independent of $P_u$.

At strong turbulent rotation, LP16's $C_P$ measures the product of intrinsic-polarization correlations and the depth over which their Faraday phases remain coherent. The direction spectrum removes amplitude weighting but requires the conditional calculation developed in Section~\ref{sec:phase_mapping}; applying a Gaussian-pair formula directly to $C_P$ can give an incorrect coherent-core amplitude. Neither normalized measure has a universal advantage in every range of wavelength and scale. The LP16 frequency variance offers a complementary way to identify the rotation regime, while the directional spectrum can reveal transverse break scales without differentiating spectral data. In the pure strong-mean limit, the LP16 spatial and interferometric spectra recover the LOS-selected emissivity fluctuations directly. Directional spectra and gradients characterize the corresponding variations of the polarization direction and have additional logarithmic behavior under the Gaussian-emissivity model.

The LP12 synchrotron-intensity spectrum supplies a separate constraint on the emitting field, because it has no Faraday phase. For a foreground screen, the polarization-direction factorization has already been studied by \citet{Melnichenka2026}. In the internally mixed case that factorization is replaced by the conditional statistics derived here. These methods consequently provide complementary constraints on the source, the Faraday-rotating plasma, and the geometry connecting them.\par

\subsection{Distinguishing emitting and Faraday-rotating regions}\label{sec:geometry}
The complementary sensitivity of the above measures can be used to establish whether synchrotron emission and Faraday rotation originate predominantly in the same volume. Consider a mixed emitting and rotating region, with emergent polarization $P_m$, behind a non-emitting Faraday screen of depth $\Phi_f(\X)$. Then
\begin{align}
 P_{\rm obs}&=P_m e^{2ix\Phi_f},\nonumber\\
 |P_{\rm obs}|&=|P_m|,\qquad u_{\rm obs}=u_m e^{2ix\Phi_f}.
 \label{eq:geometry_screen}
\end{align}
For a purely emitting source, $P_m=P_0$ after removing the intrinsic synchrotron-frequency dependence. An ideal resolved screen therefore changes the direction spectrum and the normalized polarization gradient without changing the polarized amplitude or its variance. Internal rotation generally changes both through interference of radiation from different depths. The LP16 wavelength derivative helps separate these effects:
\begin{equation}
 \frac{\partial_x P_{\rm obs}}{P_{\rm obs}}=
 \frac{\partial_x P_m}{P_m}+2i\Phi_f.
 \label{eq:geometry_logderivative}
\end{equation}
Its real part measures the wavelength variation of $\ln|P_m|$, while half its imaginary part measures the total angle rotation rate. The LP12 total-intensity spectrum provides a Faraday-independent reference for the emitting turbulence.

If the screen and mixed region are statistically independent, their directional correlations multiply:
\begin{equation}
 \xi_{u,\rm obs}=\xi_{u,m}\avg{e^{2ix\Delta\Phi_f}},
 \label{eq:geometry_factor}
\end{equation}
where $\Delta\Phi_f=\Phi_f(\X)-\Phi_f(\X+\R)$. For zero-mean Gaussian increments the last factor is $\exp[-2x^2\avg{(\Delta\Phi_f)^2}]$. Thus the mixed-medium theory, constrained by the amplitude and LP16 statistics, can be used to estimate the additional screen decorrelation wherever $\xi_{u,m}$ is measurable. The two contributions are not generally additive in the spectrum. At weak rotation, angle statistics alone measure $q_m+\Phi_f$ (Appendix~\ref{app:weak}), and cannot uniquely distinguish the two regions.

An additional test compares their projected magnetic-field orientations. Form the cross-frequency normalized polarization $v_{21}=u(\X,x_2)u^*(\X,x_1)$ and define
\begin{equation}
 \boldsymbol{g}_{F,21}=
 \frac{\Ima[v_{21}^*\nabla v_{21}]}{2(x_2-x_1)}.
 \label{eq:geometry_cyclic}
\end{equation}
This avoids differentiating a wrapped polarization angle. For a pure screen it equals $\nabla\Phi_f$ at every wavelength pair. Internal rotation instead gives $\nabla q_m+(x_1+x_2)\nabla s_m+\cdots$ in the weak limit. The orientation of $\boldsymbol{g}_{F,21}$ can be compared with that of the synchrotron-intensity gradient using $\avg{\cos 2(\theta_F-\theta_I)}$. When anisotropic turbulence makes both gradients preferentially perpendicular to their local projected magnetic fields \citep{LazarianEtAl2017SIG,LazarianYuen2018}, persistent misalignment and wavelength-pair independence favor a separate rotating region. Agreement does not establish co-spatiality, since distinct regions can share a field orientation, and density fluctuations or different LOS weights can also cause misalignment.

The strongest evidence for internal mixing is therefore a joint wavelength dependence of polarized amplitude, direction correlations, and the angle-rotation gradient. Directional changes without amplitude changes and with a wavelength-independent $\boldsymbol{g}_{F,21}$ favor foreground rotation. These tests require matched angular resolution, narrow channels, and adequate polarized signal, since unresolved screen fluctuations can also depolarize the emission.

\section{Discussion}\label{sec:discussion}
\subsection{Why the directional spectrum responds to Faraday turbulence}
Normalization removes polarized-brightness weighting at each measured sky position. Propagation can rotate the polarization without a corresponding amplitude change. It can therefore leave a much clearer imprint in directional maps than in the ordinary polarization spectrum. The separated-screen calculations of \citet{Melnichenka2026} demonstrate this advantage at moderate rotation. For a mixed volume, the conditional theory identifies an additional physical effect: differential Faraday rotation shifts the relative angle of two otherwise similar integrated signals, and also changes the relative phases of radiation from different depths. These effects need not produce the same direction correlation at a given ordinary coherence. The logarithmic behavior of a Gaussian pair is associated with the latter mechanism, while the former can rotate a signal without changing its magnitude.

The exact result in Eq.~(\ref{eq:conditional_main}) demonstrates that the departure from joint Gaussian polarization statistics can occur even for Gaussian Faraday-density fluctuations. The Brownian example further shows that the deviation can be sensitive to the balance of intrinsic and Faraday fluctuations. For a long-correlated emissivity field the conditional variances must also be retained. A theory based only on $\rho_P$ loses this information. It follows that the direction correlation potentially probes higher-order Faraday statistics, including intermittency, in addition to the ordinary Faraday coherence depth.

The coherence depth still describes the ordinary polarization and provides useful guidance to the directional spectrum under the conditions specified in Section~\ref{sec:branches}. Under strong random rotation, $\ell$ sets the overall depolarization, while $w_\phi(R)$ describes the spatial coherence. Normalization suppresses the overall amplitude variation but does not remove the spatially varying phase. Under weak rotation the leading response in a coherent uniform emitter is proportional to $\lambda^4 P_\phi(K,0)$ in the direction spectrum, with no saturation of its spectral index at $m_\phi=1$ (Appendix~\ref{app:weak}). The normalized tangential gradient gives additional weight to the small scales where this response may be measured.

The relationship to the LP16 complex-polarization and wavelength-derivative statistics is established in Section~\ref{sec:lp16compare}, including their complementary dependence on intrinsic emission and Faraday rotation. The separated-screen test of \citet{Melnichenka2026} and the mixed-volume theory considered here describe different geometries. In the former the foreground rotation multiplies the entire polarization by a single phase. In the latter each emitting depth contributes a different phase, and the conditional calculation is needed to obtain polarization-direction correlations.\par

\subsection{Observational implications}
The polarization-direction spectrum is obtained by Fourier transforming $u$ or by measuring its spatial correlation, both from ordinary Stokes maps. It retains the $\pi$ periodicity of the polarization angle and does not require angle unwrapping. The appropriate interpretation depends on the rotation regime: weak internal rotation samples the effective Faraday-depth fluctuations and cross terms; strong stochastic rotation selects the Faraday coherence window; strong mean rotation samples a source LOS Fourier wavenumber. For quantitative comparisons, the wavelength-dependent transition scales should be resolved and the polarized signal should have adequate signal-to-noise for normalization. The conditional calculation relates directional correlations to the statistical distribution of Faraday-depth histories. For finite angular resolution, noise and beam convolution must be accounted for before normalization. The independent Gaussian-emissivity model can be evaluated directly from Faraday histories. Tests with physical synchrotron emissivities should determine how correlations with the rotating medium modify the predictions.

\subsection{Synergy of Techniques}

The analysis of polarized radiation provides a powerful
way of exploring turbulence in astrophysical media. The
different techniques, both discussed in LP16 and in this paper, provide complementary ways of extracting the
information about turbulence in terms of synchrotron
emission and Faraday rotation. The techniques have different
sensitivity to the two processes above. Therefore,
combining them, it is possible to extract the statistics
of both magnetic field and fluctuating electron density.

The techniques provide different predictions for the
cases when the Faraday rotation and synchrotron polarization
originate at the same region or in different regions.
This opens ways for detailed studies of the structure
of turbulence distribution in space.

\section{Summary}\label{sec:summary}
Our principal results are as follows.
\begin{enumerate}
\item For independent circular Gaussian polarized emissivity we derived an exact correlation of polarization directions at arbitrary Faraday rotation strength. The ordinary Gaussian-pair formula is applied at fixed Faraday-depth histories, followed by averaging over those histories. This accommodates non-Gaussian Faraday fluctuations and explains why even Gaussian Faraday density does not guarantee joint Gaussianity of the observed polarization.
\item In a deep volume with short-correlated Gaussian emissivity, differential Faraday rotation has two effects: a relative rotation of the integrated signals and a change in interference among emitting depths. The leading logarithmic decorrelation is reduced by up to a factor of three in the Brownian limit, with the coefficient depending on the intrinsic and Faraday contributions. For finite emissivity correlation lengths the full conditional covariance is required.
\item Weak internal rotation measures an emission- and orientation-weighted Faraday depth, with a possible $\lambda^2$ intrinsic--Faraday cross term. For a uniform coherent emitter in a thick slab the isolated Faraday direction spectrum is $P_{u,F}\propto\lambda^4K^{-3-m_\phi}$ throughout $0<m_\phi<2$. The gradient spectrum retains this sensitivity, including for steep Faraday-density spectra, without a Fourier logarithm at $m_\phi=1$.
\item Strong stochastic rotation gives a physical coherence depth for ordinary polarization. For $1<m_\phi<2$ we obtained an additional shallow-depth interval $W_\phi\propto R^{-2/m_\phi}$ and a wavelength-dependent break $K_a\propto\lambda^{4/(2+m_\phi)}$. We corrected the near-surface normalization of the depth window. Directional spectral branches additionally require the low-coherence statistical conditions and sufficiently separated scales. The exact conditional result provides the basis for finite-range spectra.
\item Pure strong mean rotation selects the intrinsic polarized-emissivity spectrum at $k_\parallel=2\lambda^2\bar\phi$, rather than an emitting layer of fixed depth. The corrected spatial and interferometric asymptotes recover the valid LP16 variance $\avg{|P|^2}\propto\lambda^{-2(1+m)}$ in the source-dominated range. The wavelength-dependent break can probe anisotropy of the three-dimensional polarized emissivity. The full directional correlation and finite-beam integrals are useful where the gradient asymptotes converge slowly.
\item The polarization-direction spectrum and the normalized tangential gradient, $P_{\nabla u}=K^2P_u$, need only a single Stokes map at each wavelength. The sensitivity to differential Faraday rotation complements earlier synchrotron-intensity, polarization and wavelength-derivative diagnostics. Section~\ref{sec:lp16compare} shows how the two normalized measures complement LP16 polarization-spatial and wavelength-derivative statistics. Multiple wavelengths allow the rotating contribution and moving spectral boundaries to be investigated without differentiating noisy spectral data.
\item The combination of these techniques opens the way for detailed studies of the statistics and spatial distribution of turbulence.
\end{enumerate}

\begin{acknowledgments}
A.L. acknowledges the support of NSF grant AST 2307840. 
\end{acknowledgments}

\appendix
\section{Weak internal Faraday rotation}\label{app:weak}
For $x=\lambda^2$ sufficiently small and $P_0\ne0$, Eq.~(\ref{eq:Pmixed}) expands as
\begin{equation}
 P=P_0+2ixM_1-2x^2M_2+O(x^3),\quad
 M_n\equiv\int_0^L dz\,P_i\Phi^n.
 \label{eq:weakP}
\end{equation}
Here $P_0=\int_0^L P_i\,dz$. The complex-logarithm derivation and its angular average are given in Appendix~\ref{app:math:weak}. Expansion of the phase $\chi=\tfrac12\arg P$ yields
\begin{align}
 \chi(\X,x)&=\chi_0(\X)+xq(\X)+x^2s(\X)+O(x^3),\label{eq:weakchi}\\
 q&=\Rea(M_1/P_0),\nonumber\\
 s&=\Ima[(M_1/P_0)^2-M_2/P_0].\label{eq:qs}
\end{align}
The first coefficient is the emissivity- and angle-weighted internal Faraday depth:
\begin{align}
 q(\X)&=\int_0^L ds\,W(\X,s)\phi(\X,s),\label{eq:qweight}\\
 W(\X,s)&=\frac{\displaystyle\int_s^L dz\,\epsilon(\X,z)
 \cos\{2[\psi(\X,z)-\chi_0(\X)]\}}
 {\displaystyle\int_0^L dz\,\epsilon(\X,z)
 \cos\{2[\psi(\X,z)-\chi_0(\X)]\}}.\label{eq:Wweak}
\end{align}
For a uniform slab with coherent intrinsic angle, $W=1-s/L$ and $\int_0^L W^2ds=L/3$. This difference from a uniform LOS weight changes the normalization of the projected effective-Faraday variance, not its inertial-range exponent when the window is smooth.

For the difference between two LOSs, $\Delta\chi=\Delta\chi_0+x\Delta q+x^2\Delta s+O(x^3)$. The direction structure function expanded without a small-angle approximation is
\begin{equation}
\begin{split}
 D_u={}&D_{u,0}+4x\avg{\Delta q\sin2\Delta\chi_0}\\
 &+4x^2\avg{(\Delta q)^2\cos2\Delta\chi_0}\\
 &+4x^2\avg{\Delta s\sin2\Delta\chi_0}+O(x^3).
\end{split}
\label{eq:Duweak_exact}
\end{equation}
For small intrinsic angle differences this becomes
\begin{equation}
\begin{split}
 D_u\simeq&\ 4D_{\chi_0}+8\lambda^2C_{\chi_0q}
 +4\lambda^4D_q\\
 &+8\lambda^4C_{\chi_0s},
\end{split}
\label{eq:Duweak}
\end{equation}
where $D_q=\avg{(\Delta q)^2}$, $C_{\chi_0q}=\avg{\Delta\chi_0\Delta q}$, and $C_{\chi_0s}=\avg{\Delta\chi_0\Delta s}$. The $\lambda^2$ term represents intrinsic--Faraday correlations. The $\lambda^4$ coefficient is not purely a Faraday variance in general. These ensemble expansions require finite moments and a sufficiently small rotation on the emitting lines of sight. In a zero-mean Gaussian source, the distribution of $P_0$ extends to small amplitudes and the moments of $M_n/P_0$ need not be finite. In that case Eq.~(\ref{eq:conditional_main}) provides the appropriate ensemble calculation, rather than termwise averaging of a singular phase expansion.

For a coherent uniform emitter the weight $W(z)=1-z/L$ is deterministic and $q=\int W\phi\,dz$. In the isotropic thick-volume inertial range $KL\gg1$, direct Fourier projection gives
\begin{equation}
 P_q(K)=\int\frac{dk_z}{2\pi}|\widetilde W(k_z)|^2
 P_\phi(K,k_z)\simeq\frac L3 P_\phi(K,0).
 \label{eq:weakproj}
\end{equation}
For $P_\phi(k)\propto k^{-3-m_\phi}$, the resulting direction spectrum is
\begin{equation}
 P_{u,F}(K)=4\lambda^4P_q(K)\propto
 \lambda^4 K^{-3-m_\phi},\qquad 0<m_\phi<2.
 \label{eq:weakscaling}
\end{equation}
For $0<m_\phi<1$, $D_q\propto R^{1+m_\phi}$. At $m_\phi=1$, $D_q\propto R^2\ln(r_\phi/R)$, and for $m_\phi>1$ the leading structure function is proportional to $R^2$. In the latter case the nonanalytic subleading term, not the leading quadratic term, fixes the inertial-range spectrum. For a varying intrinsic angle and stochastic depth weight one must retain the full weighted and cross terms; the uniform-source spectral result is not a general formula for $P_i$ with arbitrary spatial statistics. For a coherent emitter the small differential phase criterion is $\Theta_F(R)=\lambda^2[D_q(R)]^{1/2}\ll1$. The phase expansion must also be valid over the contributing emitting depths. A fit $P_u=A_0(K)+\lambda^2 A_2(K)+\lambda^4 A_4(K)$ separates the orders where the expansion exists, but not the two parts of $A_4$.

\section{Mathematical derivations of the principal relations}\label{app:math}
This appendix derives the main mathematical relations and keeps the assumptions on emissivity and Faraday density separate.

\subsection{Gaussian Faraday averaging and finite-depth limits}\label{app:math:window}
Set $x=\lambda^2$, $\Phi=\bar\phi z+\delta\Phi$ and $Y=\delta\Phi(\X,z_1)-\delta\Phi(\X+\R,z_2)$. If $Y$ is a zero-mean Gaussian variable with $D_{\Delta\Phi}=\avg{Y^2}/2$, then
\begin{equation}
 \avg{e^{2ixY}}=e^{-(2x)^2\avg{Y^2}/2}=e^{-4x^2D_{\Delta\Phi}}.
 \label{eq:math:gaussianphase}
\end{equation}
This gives Eq.~(\ref{eq:LP16_CP_exact}) independently of Gaussianity of the emissivity, when emissivity and Faraday density are statistically independent.

With $s=z_1-z_2$ and $t=(z_1+z_2)/2$ the Jacobian is unity and $|s|\le2\min(t,L-t)$. Near $s=0$, expand $D_{\Delta\Phi}=D^+_{\Delta\Phi}+\Lambda_-s^2/4+o(s^2)$. Only if the emitting depths are sufficiently far from both faces and $\xi_i(R,s)$ varies slowly over the longitudinal coherence width can one extend $s$ over the whole line. The integral is
\begin{equation}
 \int_{-\infty}^{\infty} ds\,e^{-x^2\Lambda_-s^2+2ix\bar\phi s}
 =\frac{\sqrt\pi}{x\sqrt{\Lambda_-}}e^{-\bar\phi^2/\Lambda_-}.
 \label{eq:math:complete-square}
\end{equation}
For negligible mean rotation this gives Eq.~(\ref{eq:Wphi_def}). For a homogeneous isotropic density, direct differentiation of the variance at equal depths gives
\begin{equation}
 \Lambda_-(R,t)=C_\phi(t)-C_\phi(\sqrt{t^2+R^2})+2C_\phi(R).
 \label{eq:lambda_exact}
\end{equation}
If the quadratic expansion is accurate but the faces cannot be ignored, the zero-mean longitudinal Gaussian integral is multiplied by $\operatorname{erf}[2x\sqrt{\Lambda_-}\min(t,L-t)]$ wherever $\Lambda_->0$. Close to a face or when the quadratic expansion fails, Eq.~(\ref{eq:LP16_CP_exact}) is the reference integral. At $R=0$, $D^+=0$ and $\Lambda_-=2\sigma_\phi^2$, so $\xi_P(0)\simeq\sqrt\pi\ell L\xi_i(0,0)$. At finite $R$, the mean exponential can be neglected only when $\bar\phi^2\ll\Lambda_-$.

For the isotropic Faraday-density example in the main text the equal-depth variance is
\begin{equation}
 D^+_{\Delta\Phi}(R,t)=2\int_0^t ds\,(t-s)
 [C_\phi(s)-C_\phi(\sqrt{s^2+R^2})].
 \label{eq:math:equaldepth}
\end{equation}
The factor of two follows from integrating the covariance of the difference of the two densities and dividing the variance by two. For $R\ll t\ll r_\phi$, the difference in brackets is $\sigma_\phi^2 r_\phi^{-m_\phi}[(s^2+R^2)^{m_\phi/2}-s^{m_\phi}]$. For $s\gg R$, its leading term is $(m_\phi/2)\sigma_\phi^2r_\phi^{-m_\phi}R^2 s^{m_\phi-2}$.

If $m_\phi<1$, the integral is dominated by $s\sim R$. Then $D^+\sim\sigma_\phi^2t R^{1+m_\phi}/r_\phi^{m_\phi}$ for $t\gg R$. If $m_\phi>1$ and $R\ll t\ll r_\phi$, the upper part of the integral contributes and
\begin{equation}
 D^+\sim\sigma_\phi^2R^2(t/r_\phi)^{m_\phi}.
 \label{eq:math:middledepth}
\end{equation}
For $t\gg r_\phi$ in the steep case, $D^+\sim\sigma_\phi^2tR^2/r_\phi$ provided the covariance decays sufficiently rapidly. More generally define
\begin{equation}
 I_\phi(R)=\int_0^\infty ds\,[C_\phi(s)-C_\phi(\sqrt{s^2+R^2})].
 \label{eq:depth_I}
\end{equation}
For depths much greater than the longitudinal correlation length of the density difference, $D^+\simeq2t I_\phi(R)$. At small $R$ and negligible intrinsic variation, the prefactor in Eq.~(\ref{eq:Wphi_def}) approaches unity. In the deep, low-coherence branch this gives $W_\phi\simeq[8x^2I_\phi(R)]^{-1}$, or $(1-e^{-8x^2I_\phi L})/(8x^2 I_\phi)$ when the finite upper depth must be retained within the deep approximation. For $m_\phi<1$, the leading $I_\phi\propto R^{1+m_\phi}$ receives an analytic $R^2$ contribution. For $m_\phi>1$,
\begin{equation}
 I_\phi(R)\simeq-\frac{R^2}{2}\int_0^\infty\frac{C_\phi'(s)}s\,ds,
 \label{eq:depth_steep_coeff}
\end{equation}
with a nonanalytic correction of order $R^{1+m_\phi}$. Thus for a smooth outer-scale model with the specified small-scale power, the relative corrections can decrease only as $(R/r_\phi)^{|1-m_\phi|}$. Their coefficients depend on the full covariance. The depth coefficient therefore approaches its limiting value slowly for indices close to unity. The marginal result is $I_\phi\sim\sigma_\phi^2R^2\ln(r_\phi/R)/(2r_\phi)$. Finally, for $t\ll R\ll r_\phi$, $D^+\sim\sigma_\phi^2t^2(R/r_\phi)^{m_\phi}$. At $m_\phi=1$ and $t\gg R$ the variance has the additional factor $\ln[\min(t,r_\phi)/R]$ relative to $\sigma_\phi^2tR^2/r_\phi$. Setting $4x^2D^+\sim1$ gives Eqs.~(\ref{eq:Wphi_branches}) and (\ref{eq:Wphi_steep}). The steep intermediate depth becomes $r_\phi$ at $R_b\sim\ell$. Setting the shallow depth to $R$ gives $R_a/r_\phi\sim(\ell/r_\phi)^{2/(2+m_\phi)}$. Setting the deep depth to $L$ gives $R_c$ in Eq.~(\ref{eq:Rc_strong}). At $m_\phi=1$ one should solve the depth condition with the logarithm retained, not extrapolate the pure powers.

For widely separated lines of sight only depths of order $\ell$ contribute. Over these depths $\Phi_a(z)\simeq z\phi_a(0)$ and the near-surface covariance is $c\sigma_\phi^2$ between the two Faraday densities. Neglecting mean rotation and variations of the source along these short depths,
\begin{equation}
 \frac{\xi_P(R)}{\xi_i(R,0)}\simeq\int_0^\infty dz_1\int_0^\infty dz_2
 e^{-(z_1^2+z_2^2-2cz_1z_2)/\ell^2}.
 \label{eq:math:quadrant}
\end{equation}
The Gaussian integral over the physical quadrant is $\ell^2[\pi/2+\arcsin c]/[2\sqrt{1-c^2}]$. This follows either by polar coordinates or by integrating a bivariate Gaussian over its positive quadrant. Dividing by $\sqrt\pi\ell L$ proves Eq.~(\ref{eq:plateau}). At $c=0$ each one-sided integral is $\sqrt\pi\ell/2$. Extending the difference-variable integral to the whole line before applying the depth boundary misses the factor of two. The approximation for nonzero $c$ also requires that the resulting contributing depths are shorter than $r_\phi$ and $L$.

\subsection{The normalized directions of a correlated complex Gaussian pair}\label{app:math:phase}
Let $Z_1,Z_2$ be zero-mean, unit-variance, circular complex Gaussian variables with
\begin{equation}
 \langle Z_1Z_2^*\rangle=\rho,
 \qquad \langle Z_1Z_2\rangle=0,
 \qquad |\rho|<1.
 \label{eq:math:cov}
\end{equation}
Their covariance matrix has determinant $1-|\rho|^2$. Rotating one variable makes the covariance real and positive, $r=|\rho|$, while the desired directional expectation acquires the phase factor $\rho/r$. For real $r$, the joint density is
\begin{equation}
 p(z_1,z_2)=\frac{1}{\pi^2(1-r^2)}
 \exp\!\left[-\frac{|z_1|^2+|z_2|^2-2r\Rea(z_1z_2^*)}
 {1-r^2}\right].
 \label{eq:math:jointpdf}
\end{equation}
Write $z_1=a e^{i\alpha}$, $z_2=b e^{i\beta}$ with $a,b>0$. The expectation of $u_1u_2^*=e^{i(\alpha-\beta)}$ is obtained using
\begin{equation}
 \int_0^{2\pi}d\theta\,e^{i\theta}e^{t\cos\theta}
 =2\pi I_1(t),
 \label{eq:math:angularBessel}
\end{equation}
where $I_1$ is the modified Bessel function. The remaining radii give
\begin{equation}
 \langle u_1u_2^*\rangle_r=
 \frac{4}{1-r^2}\int_0^\infty a\,da\int_0^\infty b\,db\,
 e^{-(a^2+b^2)/(1-r^2)}
 I_1\!\left(\frac{2rab}{1-r^2}\right).
 \label{eq:math:radial}
\end{equation}
Now substitute the absolutely convergent series
\begin{equation}
 I_1(2t)=\sum_{n=0}^\infty
 \frac{t^{2n+1}}{n!\,(n+1)!}
 \label{eq:math:I1series}
\end{equation}
and integrate termwise with
\begin{equation}
 \int_0^\infty da\,a^{2n+2}e^{-a^2/h}
 =\frac12 h^{n+3/2}\Gamma\!\left(n+\frac32\right).
 \label{eq:math:radint}
\end{equation}
With $h=1-r^2$, the result is
\begin{align}
 \langle u_1u_2^*\rangle_r
 &=r(1-r^2)\sum_{n=0}^\infty
 \frac{\Gamma(n+3/2)^2}{n!\,(n+1)!}\,r^{2n}\nonumber\\
 &=\frac\pi4 r(1-r^2)
 {}_2F_1\!\left(\frac32,\frac32;2;r^2\right)\nonumber\\
 &=\frac\pi4 r\,
 {}_2F_1\!\left(\frac12,\frac12;2;r^2\right).
 \label{eq:math:hypergeom}
\end{align}
The final equality is Euler's hypergeometric transformation. Restoring the complex phase replaces $r$ by $\rho$ outside the hypergeometric function while leaving $r^2=|\rho|^2$ in its argument. This is Eq.~(\ref{eq:phase_gaussian_exact}). In particular, at small $r$, $\langle u_1u_2^*\rangle=(\pi/4)\rho+O(\rho|\rho|^2)$, rather than $\rho$.

For the nearly coherent limit take real $\rho=r$ and $\varepsilon=1-r^2$. A convenient equivalent form is
\begin{equation}
 \langle u_1u_2^*\rangle
 =\frac{E(1-\varepsilon)-\varepsilon K(1-\varepsilon)}
 {\sqrt{1-\varepsilon}},
 \label{eq:math:elliptic}
\end{equation}
where $K(z)$ and $E(z)$ are complete elliptic integrals with parameter $z$. Their standard expansions at $\varepsilon\downarrow0$ are
\begin{align}
 K(1-\varepsilon)&=\ln(4/\sqrt\varepsilon)
 +O(\varepsilon\ln\varepsilon),\nonumber\\
 E(1-\varepsilon)&=1+\frac\varepsilon2
 \left[\ln(4/\sqrt\varepsilon)-\frac12\right]
 +O(\varepsilon^2\ln\varepsilon).
 \label{eq:math:elliptexp}
\end{align}
It follows that
\begin{equation}
 D_u=2(1-\langle u_1u_2^*\rangle)
 =\frac\varepsilon2\left[\ln\!\left(\frac{16}{\varepsilon}\right)-1\right]
 +O(\varepsilon^2\ln\varepsilon).
 \label{eq:math:Du-log}
\end{equation}
This is the logarithmic limit of the Gaussian-pair function. Under random Faraday rotation, its argument must be the conditional correlation in Eq.~(\ref{eq:conditional_main}) before averaging.

\subsection{Conditional averaging and a Brownian differential phase}\label{app:math:conditional}
For fixed $\Phi$, insert Eq.~(\ref{eq:Pmixed}) in $\avg{P_aP_b^*}_{P_i|\Phi}$ and use independence to obtain Eq.~(\ref{eq:conditional_cov}). The pseudo-covariance $\avg{P_aP_b}_{P_i|\Phi}$ vanishes since $\avg{P_iP_i}=0$. Thus the direction average of the conditional pair is exactly Eq.~(\ref{eq:phase_gaussian_exact}) evaluated at $\rho_c$. The law of total expectation gives Eq.~(\ref{eq:conditional_main}). For a general nonzero emission mean or noncircular Gaussian field the conditional mean or pseudo-covariance need not vanish; the two-variable law must then include them.

In the short-correlated model the modulus squared of the conditional coherence has the expansion $|\rho_c|^2=1-a-V_\theta+o(a+V_\theta)$, while its phase is $\bar\theta$ to first order. Multiplying the expansion of Eq.~(\ref{eq:phase_gaussian_exact}) by $\cos\bar\theta$ for its real part proves Eq.~(\ref{eq:conditional_near}). The ordinary decorrelation is $\varepsilon\simeq a+\avg{\overline{\theta^2}}$, where $\overline{\theta^2}=L^{-1}\int\theta^2dz=\bar\theta^2+V_\theta$. This decomposition is the reason the original unconditional Gaussian result cannot in general be used.

For Brownian motion $\avg{W(t)W(t')}=\min(t,t')$ implies $\avg{\int_0^1 W(t)^2dt}=1/2$ and $\avg{\bar W^2}=\int_0^1\int_0^1\min(t,t')dt dt'=1/3$. Subtracting gives $\avg V=1/6$. The exact ordinary coherence for $a=0$ is the average of Eq.~(\ref{eq:white_conditional}):
\begin{equation}
 \rho_P=\int_0^1 dt\,e^{-s^2t/2}=\frac{1-e^{-s^2/2}}{s^2/2}.
 \label{eq:brownian_rho}
\end{equation}
Hence $\varepsilon\simeq s^2/2$ for $s\to0$. Substituting $V_\theta=s^2 V$ into Eq.~(\ref{eq:conditional_near}) gives
\begin{equation}
 1-\xi_u=\frac{\varepsilon}{12}\left[
 \ln\frac{16}{\varepsilon}+3-\ln 2-6\avg{V\ln V}\right]+o(\varepsilon).
 \label{eq:math:brownian}
\end{equation}
The centered Brownian covariance has eigenfunctions proportional to $\cos(n\pi t)$, $n\ge1$, with eigenvalues $(n\pi)^{-2}$, obtained by solving the inverse one-dimensional Laplacian with Neumann boundary conditions and removing the constant mode. Thus $V$ is distributed as $\sum_{n=1}^\infty Z_n^2/(n\pi)^2$, where $Z_n$ are independent standard normal variables. Its Laplace transform is
\begin{equation}
 H(t)=\avg{e^{-tV}}=\left(\frac{\sqrt{2t}}{\sinh\sqrt{2t}}\right)^{1/2}.
\end{equation}
Using $\ln v=\int_0^\infty(e^{-t}-e^{-vt})dt/t$ gives
\begin{equation}
 \avg{V\ln V}=\int_0^\infty\frac{e^{-t}/6+H'(t)}{t}\,dt.
 \label{eq:math:vlogv}
\end{equation}
Equations~(\ref{eq:math:brownian}) and (\ref{eq:math:vlogv}) determine $C_B$ in Eq.~(\ref{eq:brownian_core}). Nonzero $a$ gives Eq.~(\ref{eq:mixed_core}).

\paragraph{Uniform coherent emissivity.} A different conditional approximation applies when $P_i$ is a nonrandom constant. For two equivalent sightlines with jointly Gaussian Faraday depths, let $\Phi_s=(\Phi_1+\Phi_2)/2$, $\delta\Phi=\Phi_1-\Phi_2$, and $v(z)=e^{ix\delta\Phi(z)}$. The fluctuating sum and difference are independent because their cross covariance vanishes by sightline exchange symmetry. At fixed $\delta\Phi$, write $P_1/P_i=\int dz\,e^{2ix\Phi_s(z)}v(z)$ and $P_2/P_i=\int dz\,e^{2ix\Phi_s(z)}v^*(z)$. At negligible mean rotation the common-phase kernel is real:
\begin{equation}
 G(z,z')=\exp[-2x^2\avg{(\Phi_s(z)-\Phi_s(z'))^2}].
\end{equation}
The conditional second moments of $P_1,P_2$ follow by integrating this kernel. For $\ell\ll r_\phi$ the common-phase sum collects about one stationary-phase contribution, with a random phase, per $r_\phi$ of depth, where $\phi_s=d\Phi_s/dz$ changes sign; it therefore becomes circular Gaussian for $L\gg r_\phi$, with corrections of order $r_\phi/L$. If, in addition, the conditional means and pseudo-covariances are negligible, one obtains
\begin{equation}
 \xi_u\simeq\avg{{\cal F}(\rho_{\delta})}_{\delta\Phi},\qquad
 \rho_{\delta}=\frac{\int dz\,dz'\,G(z,z')v(z)v(z')}
 {\int dz\,dz'\,G(z,z')v(z)v^*(z')}.
 \label{eq:uniform_conditional}
\end{equation}
The two conditional variances are equal here because $G$ is real and symmetric. This is the conditional approximation proposed by A. Melnichenka for a uniform emitting volume. The mean and pseudo-covariance are not identically zero: in a finite volume the common-phase integral is not Gaussian, and the front surface contributes a mean. For sufficiently deep strongly rotating volumes and differential phases varying little over the width of $G$, replacing $G$ by a longitudinal delta function with nearly constant weight recovers Eq.~(\ref{eq:white_conditional}) with $g_i=1$. Its Brownian limit gives Eq.~(\ref{eq:brownian_core}). The uniform-source approximation is distinct from the exact Gaussian-emissivity result in Section~\ref{sec:phase_mapping}.

\subsection{Fourier transforms and gradient spectra}\label{app:math:hankel}
For isotropic maps $P_u(K)=2\pi\int_0^\infty R\,dR\,J_0(KR)\Xi_u(R)$. For $\Xi_u=A R^{-\gamma}$, the substitution $t=KR$ gives
\begin{equation}
 P_u=2\pi A\,2^{1-\gamma}
 \frac{\Gamma(1-\gamma/2)}{\Gamma(\gamma/2)}K^{\gamma-2}.
 \label{eq:math:hankelpower}
\end{equation}
The lower end of the ordinary improper integral converges for $\gamma<2$, and the upper end converges by oscillation for $\gamma>1/2$. For $\gamma=2$, integration starting at $R_c$ gives $\ln(K_c/K)$ plus finite terms when $K_c/K$ is large. For a finite interval replace the power-law transform by the corresponding integral over its limits, as in Eq.~(\ref{eq:finite_window_hankel}). This derivation concerns the correlation of directions itself; a similar power of $W_\phi$ gives these slopes only with the low-coherence conditions of Eq.~(\ref{eq:phase_lowrho}).

For the small-$R$ core, write the regularized two-dimensional transform of $R^p$ at $K>0$ as $A(p)K^{-2-p}$, with
\begin{equation}
 A(p)=2^{p+2}\pi\frac{\Gamma(1+p/2)}{\Gamma(-p/2)}.
 \label{eq:math:powertransform}
\end{equation}
Analytic continuation, or differentiation with a smooth cutoff, gives the high-$K$ coefficient of a local nonanalytic term. The transform of $R^p\ln(R_0/R)$ is $K^{-2-p}[A(p)\ln(KR_0)-A'(p)]$. For $0<p<2$ this implies the logarithmic correction in Eq.~(\ref{eq:tangential_core}). For $p=2$, $A(2)=0$ but $A'(2)=8\pi$, so $R^2\ln(R_0/R)$ gives $K^{-4}$ with no additional Fourier logarithm. A term $R^2\ln^2(R_0/R)$ instead produces a logarithm in the $K^{-4}$ spectrum. Pure even powers contribute only distributions at $K=0$ in this local asymptotic calculation. Since $\Xi_u=\Xi_u(0)-D_u/2$ at nonzero modes, the sign of the transform of $D_u$ must be reversed and divided by two. For example, Eq.~(\ref{eq:core_brownian_spatial}) yields for $0<\beta<2$ the explicit asymptotic amplitude
\begin{equation}
\begin{split}
 P_{\nabla u}(K)\simeq&-\frac{A(\beta)}{12R_0^\beta}K^{-\beta}\bigg[\beta\ln(KR_0)\\
 &+\ln16+C_B-\beta\frac{A'(\beta)}{A(\beta)}\bigg].
\end{split}\label{eq:core_brownian_spectrum}
\end{equation}
This is a Faraday-dominated Brownian result for a local power of the ordinary decorrelation. In the endpoint case $\beta=2$, taking the transform of the real-space expression before taking the limit gives $P_{\nabla u}\simeq(4\pi/3)R_0^{-2}K^{-2}$ and $\avg{|\nabla u_b|^2}\simeq (2/3)R_0^{-2}\ln(R_0/b)$ at leading order. The finite-depth spectrum should be used when $KR_0$ is not large. Finally $\widetilde{\nabla u}=i\K\widetilde u$, giving $P_{\nabla u}=K^2P_u$ exactly. Smoothing with a Gaussian of rms width $b$ gives $\avg{|\nabla u_b|^2}=\int d^2\K\,K^2 P_u(\K)e^{-b^2 K^2}/(2\pi)^2$, which proves the beam scalings in Section~\ref{sec:tangential} when the corresponding high-$K$ spectrum dominates.

\subsection{Complex-logarithm derivation of the weak-rotation terms}\label{app:math:weak}
Write $A=M_1/P_0$ and $B=M_2/P_0$ in Eq.~(\ref{eq:weakP}). Since $\chi=\frac12\Im\ln P$, expansion about $P_0\ne0$ gives
\begin{align}
 \ln\frac{P}{P_0}
 &=\ln(1+2ixA-2x^2B+O(x^3))\nonumber\\
 &=2ixA+2x^2(A^2-B)+O(x^3),\nonumber\\
 \chi-\chi_0
 &=x\Rea A+x^2\Ima(A^2-B)+O(x^3).
 \label{eq:math:complexlog}
\end{align}
This proves the coefficients $q$ and $s$ in Eq.~(\ref{eq:qs}). To obtain the depth weight, write $P_i(z)=\epsilon(z)e^{2i\psi(z)}$ and $P_0=|P_0|e^{2i\chi_0}$. Then
\begin{align}
 q&=\frac{1}{|P_0|}\int_0^L dz\,\epsilon(z)
 \Phi(z)\cos[2(\psi(z)-\chi_0)]\nonumber\\
 &=\int_0^L ds\,\phi(s)\,
 \frac{\displaystyle\int_s^L dz\,\epsilon(z)
 \cos[2(\psi(z)-\chi_0)]}
 {|P_0|},
 \label{eq:math:weightderiv}
\end{align}
where the second line interchanges the $s$ and $z$ integrals in $\Phi(z)=\int_0^z\phi(s)ds$. The equality $|P_0|=\int_0^L dz\,\epsilon(z)\cos[2(\psi(z)-\chi_0)]$ converts the last fraction to the $W(s)$ in Eq.~(\ref{eq:Wweak}).

For two sky positions let $a=2\Delta\chi_0$ and $\delta=2x\Delta q+2x^2\Delta s$. Inserting
\begin{equation}
 \cos(a+\delta)=\cos a-\delta\sin a
 -\frac{\delta^2}{2}\cos a+O(\delta^3)
 \label{eq:math:cosseries}
\end{equation}
into $D_u=2[1-\langle\cos(a+\delta)\rangle]$ gives Eq.~(\ref{eq:Duweak_exact}) without an assumption about the size of $\Delta\chi_0$. If $|\Delta\chi_0|\ll1$, use $\sin 2\Delta\chi_0\simeq2\Delta\chi_0$ and $\cos 2\Delta\chi_0\simeq1$ to obtain Eq.~(\ref{eq:Duweak}). For the uniform coherent source, $q=\int W(z)\phi(\X,z)dz$. Fourier expansion of the Faraday density gives Eq.~(\ref{eq:weakproj}). The integral of $|\widetilde W(k_z)|^2/(2\pi)$ is $\int_0^L W^2dz=L/3$. For $KL\gg1$, the source spectrum is nearly constant over the dominant range of longitudinal wavenumbers selected by the weight. Thus the thick-volume weak-direction spectrum follows with exponent $-3-m_\phi$ for the entire interval $0<m_\phi<2$. At $m_\phi=1$ the logarithm occurs in the structure function, not the Fourier spectrum. At $m_\phi>1$ the small-separation quadratic contribution to the structure function reflects the finite gradient variance, while the first nonanalytic term is of order $R^{1+m_\phi}$. Equation~(\ref{eq:math:powertransform}) explains why the leading analytic term cannot be used to infer the nonzero-$K$ spectral exponent.

\subsection{Corrected mean-phase Fourier selection and isotropic benchmark}\label{app:math:mean}
In the pure-mean limit define $k=2x\bar\phi$. Multiplication of $P(\X,x)=\int_0^L dz\,P_i(\X,z)e^{ikz}$ and its conjugate at a displaced sky position yields
\begin{align}
 \xi_P(R,x)&=\int_0^Ldz_1\int_0^Ldz_2\,
 e^{ik(z_1-z_2)}\xi_i(R,z_1-z_2)\nonumber\\
 &=\int_{-L}^Lds\,(L-|s|)e^{iks}\xi_i(R,s).
 \label{eq:math:triangular}
\end{align}
There is no stationary point of the linear phase $ks$; the integral performs Fourier cancellation rather than positive depth truncation. The exact constant-correlation counterexample is Eq.~(\ref{eq:cosexample}). When the source short-distance contribution is dominant, the triangle kernel may be replaced by $L$ over that contribution, giving Eq.~(\ref{eq:CP_mean_Fourier}).

For a source cusp $\xi_i(0,s)=\xi_i(0,0)-C_i|s|^m+o(|s|^m)$, introduce $e^{-\eta|s|}$ with $\eta\downarrow0$ to regulate the oscillatory integral:
\begin{align}
 \int_{-\infty}^\infty ds\,|s|^m e^{iks-\eta|s|}
 &=2\Gamma(1+m)\Rea(\eta-ik)^{-1-m}\nonumber\\
 &\longrightarrow-2\Gamma(1+m)
 \sin\!\left(\frac{\pi m}{2}\right)|k|^{-1-m}.
 \label{eq:math:cusp}
\end{align}
The minus sign in the cusp of $\xi_i$ gives the positive coefficient in Eq.~(\ref{eq:mean_variance}). A sharp slab contributes face terms of order $k^{-2}$, so $0<m<1$ is the interval in which the $k^{-1-m}$ cusp dominates without additional apodization.

For the interferometric relation define $S_i(\K,q)$ as in Eq.~(\ref{eq:source_spec}). Inserting this spectral representation into Eq.~(\ref{eq:math:triangular}) and Fourier transforming over $\R$ gives
\begin{align}
 S_P(\K,x)&=\int\frac{dq}{2\pi}S_i(\K,q)
 \int_{-L}^Lds\,(L-|s|)e^{i(k-q)s}\nonumber\\
 &=\int\frac{dq}{2\pi}S_i(\K,q)
 \left|\int_0^Ldz\,e^{i(k-q)z}\right|^2.
 \label{eq:math:specconv}
\end{align}
The squared window has total area
\begin{equation}
 \int_{-\infty}^\infty\frac{dp}{2\pi}|W_L(p)|^2=L.
 \label{eq:math:parsevalwindow}
\end{equation}
If $S_i(\K,q)$ varies little over $|q-k|\lesssim L^{-1}$, the convolution becomes $S_P(\K,x)\simeq LS_i(\K,k)$.

For isotropic $S_i(K,q)=A(K^2+q^2)^{-\alpha}$, $\alpha=(3+m)/2$, the two-dimensional Fourier inversion uses the identity
\begin{equation}
 \int_0^\infty K\,dK\,J_0(KR)(K^2+k^2)^{-\alpha}
 =\frac{R^{\alpha-1}|k|^{1-\alpha}}{2^{\alpha-1}\Gamma(\alpha)}
 K_{\alpha-1}(|k|R).
 \label{eq:math:bessel-transform}
\end{equation}
Dividing by its value at $R=0$ and setting $\nu=\alpha-1=(1+m)/2$ gives Eq.~(\ref{eq:rho_strong_mean}). For $0<m<1$, the small-$y$ expansion of $K_\nu(y)$ yields
\begin{equation}
 1-\rho_P(R)=
 -\frac{\Gamma(-\nu)}{2^{2\nu}\Gamma(\nu)}
 (|k|R)^{2\nu}+o(R^{2\nu}),
 \quad 0<\nu<1.
 \label{eq:math:small-bessel}
\end{equation}
Substitution in Eq.~(\ref{eq:math:Du-log}) with $\varepsilon\simeq2(1-\rho_P)$ gives $D_u\propto R^{1+m}\ln(1/R)$, followed by $P_u\propto K^{-3-m}\ln K$ in the corresponding high-$K$ range. This last mapping uses the circular complex Gaussian polarization-pair law, whereas the Fourier-selected $S_P$ and its $k$-dependent break do not.

For practical wavenumbers it is useful to retain the next term in the Bessel expansion. Let $y=2R/L_{\bar\phi}$, $\mu=1+m=2\nu$ and $c_1=2^{-\mu}\Gamma(1-\nu)/\Gamma(1+\nu)$. Then
\begin{equation}
 1-\rho_P=c_1 y^\mu-\frac{y^2}{4(1-\nu)}+\cdots.
\end{equation}
Write $d=1-\rho_P$; the Gaussian-pair expansion gives $1-\xi_u=(d/2)[\ln(8/d)-1]+O(d^2\ln d)$. The first terms after substitution are
\begin{equation}
\begin{split}
 1-\xi_u={}&\frac{c_1}{2}y^\mu
 [\ln(8/c_1)-1-\mu\ln y]\\
 &-\frac{y^2}{8(1-\nu)}
 [\ln(8/c_1)-2-\mu\ln y]+\cdots .
\end{split}
\end{equation}
The second term, although quadratic in $y$ apart from its logarithm, contributes at nonzero $K$ through that logarithm. Applying Eq.~(\ref{eq:math:powertransform}) term by term with $k=KL_{\bar\phi}/2$ gives the first two spectral orders (up to the common Fourier normalization),
\begin{equation}
\begin{split}
 P_{\nabla u}&\propto k^{-\mu}[\ln k+c_* - B_* k^{-(2-\mu)}+\cdots],\\
 c_*&=\frac{\ln(8/c_1)-1}{\mu}-\frac{A'(\mu)}{A(\mu)},\\
 B_*&=-\frac{A'(2)}{4(1-\nu)c_1 A(\mu)}.
\end{split}\label{eq:math:mean_correction}
\end{equation}
Here $A(\mu)<0$ and $B_*>0$. The extra term vanishes only as a small power as $m\to1$. Higher terms are generated by the nonlinear logarithm and the higher powers in the Bessel series; the full transform is preferable when they are appreciable. This explains why the leading strong-mean gradient power with a logarithm can be a poor quantitative approximation on a finite inertial range. It does not change the Fourier selection of ordinary polarization.

The smoothed gradient variance can be calculated without a numerical Hankel transform. In dimensionless coordinates $y=2R/L_{\bar\phi}$, let $h=2b/L_{\bar\phi}$ and $f(y)=1-{\cal F}[\rho_P(y)]$. Smoothing the direction correlation with the autocorrelation of the beam and applying its Laplacian gives
\begin{equation}
 L_{\bar\phi}^2\avg{|\nabla u_b|^2}=
 \frac{4}{h^2}\int_0^\infty dt\,e^{-t}(t-1)f(2h\sqrt t).
 \label{eq:mean_beam_integral}
\end{equation}
This follows from $G_h(y)=(4\pi h^2)^{-1}e^{-y^2/(4h^2)}$, $\nabla_y^2G_h=(y^2/(4h^4)-1/h^2)G_h$, and $t=y^2/(4h^2)$. It is equivalent to Eq.~(\ref{eq:beam_exact}) and is valid at finite $b$ for the isotropic Gaussian source model. For a term $y^p(c_{p,0}+c_{p,1}\ln y)$ in $f$, its contribution is
\begin{equation}
 2h^{p-2}\{c_{p,0}J(p)+c_{p,1}[J(p)\ln h+J'(p)]\},
 \quad J(p)=p2^p\Gamma(1+p/2).
 \label{eq:mean_beam_series}
\end{equation}
For the leading $p=\mu$ term, $c_{\mu,0}=c_1[\ln(8/c_1)-1]/2$ and $c_{\mu,1}=-\mu c_1/2$. The leading term and its constant can therefore be written as
\begin{equation}
 L_{\bar\phi}^2\avg{|\nabla u_b|^2}\simeq
 \mu c_1 J(\mu)h^{\mu-2}[\ln(1/h)+c_b],
 \label{eq:mean_beam_asymp}
\end{equation}
where $c_b=[\ln(8/c_1)-1]/\mu-J'(\mu)/J(\mu)$. The quadratic logarithmic term and higher orders contribute further corrections through Eq.~(\ref{eq:mean_beam_series}). This recovers the logarithmic beam law in Section~\ref{sec:tangential} and explains why dropping its constant can lead to a large error. Direct evaluation of Eq.~(\ref{eq:mean_beam_integral}) avoids truncating this series. 

\bibliography{references}
\end{document}